\documentclass[twocolumn]{aastex63}
\usepackage{amsmath}
\DeclareMathOperator\erf{erf}
\hypersetup{linkcolor=blue,citecolor=cyan,filecolor=green,urlcolor=magenta}

\shorttitle{A Blind Asteroid Detection Pipeline}
\shortauthors{Golovich et al.}

\begin{document}

\title{A New Blind Asteroid Detection Scheme}

\correspondingauthor{Nathan Golovich}
\email{golovich1@llnl.gov}

\author[0000-0003-2632-572X]{Nathan Golovich}
\affiliation{Lawrence Livermore National Laboratory, 7000 East Avenue, Livermore, CA 94550, USA}

\author[0000-0003-3397-7021]{Noah Lifset}
\affiliation{Lawrence Livermore National Laboratory, 7000 East Avenue, Livermore, CA 94550, USA}

\author[0000-0002-6911-1038]{Robert Armstrong}
\affiliation{Lawrence Livermore National Laboratory, 7000 East Avenue, Livermore, CA 94550, USA}

\author[0000-0003-4338-8095]{Eric Green}
\affiliation{Lawrence Livermore National Laboratory, 7000 East Avenue, Livermore, CA 94550, USA}

\author[0000-0002-8505-7094]{Michael D. Schneider}
\affiliation{Lawrence Livermore National Laboratory, 7000 East Avenue, Livermore, CA 94550, USA}

\author{Roger Pearce}
\affiliation{Lawrence Livermore National Laboratory, 7000 East Avenue, Livermore, CA 94550, USA}
\affiliation{Center for Applied Scientific Computing, 7000 East Avenue, Livermore, CA 94550, USA}

\begin{abstract}
As astronomical photometric surveys continue to tile the sky repeatedly, the potential to push detection thresholds to fainter limits increases; however, traditional digital-tracking methods cannot achieve this efficiently beyond time scales where motion is approximately linear. In this paper we prototype an optimal detection scheme that samples under a user defined prior on a parameterization of the motion space, maps these sampled trajectories to the data space, and computes an optimal signal-matched filter for computing the signal to noise ratio of trial trajectories. We demonstrate the capability of this method on a small test data set from the Dark Energy Camera. We recover the majority of asteroids expected to appear and also discover hundreds of new asteroids with only a few hours of observations. We conclude by exploring the potential for extending this scheme to larger data sets that cover larger areas of the sky over longer time baselines. 
\end{abstract}


\section{Introduction}\label{sec:intro}
The vast majority of objects that exist within the solid angle of an astronomical exposure are undetectable above the noise of the image. Typical detection schemes employ a point spread function (PSF) or other convolution-based detection method and define ``detections'' as fluctuations above some threshold, typically measured in units of the signal-to-noise ratio (SNR). The number of false positives grows as this threshold is lowered such that a threshold below $\text{SNR}=5$ is rare and below $\text{SNR}=3$ is not typically useful for large, modern cameras. The most straightforward way to improve detection sensitivity is to increase the exposure time, but moving objects exhibit trailing losses and this does not allow for the removal of cosmic rays and other imprinted correlated noise that appear differently in each exposure and add over time. A common method for increasing detection sensitivity is taking multiple exposures and stacking the images and performing object detection on the median stack of the images. This reduces the noise proportional to the square root of the number of exposures (assuming equal exposure times and similar, uncorrelated noise in each), and thus allows for fainter detection thresholds. This method works well with modern cameras which have small read noise and dark current. Faint stars and galaxies are frequently observed for hours using this method.

Methods to stack the signal of moving objects are more challenging in practice. Large scale asteroid surveys such as the Catalina Sky Survey \citep[CSS;][]{1998BAAS...30.1037L}, the Panoramic Survey Telescope and Rapid Response System \citep[Pan-STARRS;[]{2016arXiv161205560C}, and the Zwicky Transient Facility \citep[ZTF;][]{2019PASP..131a8003M} do not detect moving objects systematically with such methods. Instead, each of these surveys implement a similar method where objects are detected above an SNR threshold in single-epochs before they are linked together into orbital tracks and monitored until a confident orbital determination is made. The Pan-STARRS Moving Object Processing System \citep[MOPS;][]{2013PASP..125..357D} is an example method, which is also forming the basis for the method used by the upcoming Legacy Survey of Space and Time \citep[LSST, formerly the Large Synoptic Survey Telescope;][]{2019ApJ...873..111I}, which will utilize a MOPS-like framework and report tracklets to the Minor Planet Center \citep[MCP:\footnote{https://minorplanetcenter.net/data}][]{vo:mpcorb_web}.

On the other hand, special surveys for moving objects have been carried out in order to go deeper than these single-epoch methods allow. It is informative to first consider the angular rate of motion of various classes of solar system objects of interest. The furthest comets, asteroids and Kuiper Belt objects (KBOs) move slowly enough to permit (without trailing losses) exposures of up to $\sim$20 minutes; whereas inner solar-system asteroids in the main-belt (MBAs) and near-Earth objects (NEOs) permit exposures of $\sim$2 minutes and as short as $\sim$1 second, respectively \citep{2014ApJ...782....1S, 2015AJ....150..125H}. In order to overcome this motion, ``digital tracking'' methods have been developed. Digital tracking refers to the stacking of consecutive short exposures (short enough to avoid substantial trailing losses) along moving trajectories in order to add the signal of a moving object. This gives the benefits of tracking the motion of a moving object and taking a long exposure, with the added benefit of simpler removal of stationary objects through standard difference imaging. Furthermore, by not assuming the motion of the objects of interest, the intrinsic scatter in angular motion between moving objects may be probed generically by stacking along a fine grid in angular motion space and detecting objects on ``shifted'' median stacks (due to this methodology, this technique is also often known colloquially as ``shift-and-stack'').

\citet{1992AAS...181.0610T} first applied this method to search for KBOs, which resulted in a null detection but demonstrated the capability to search for moving objects down to faint limits without dealing with trailing losses. Later, \citet{1995ApJ...455..342C} demonstrated the power of this method, by discovering a large population of KBOs with $m_{V} \leq 28$ using the Hubble Space Telescope (HST). \citet{2004AJ....128.1364B} also used HST to conduct a search for trans-Neptunian objects (TNOs) complete at the 50\%-level down to $m_{606W}\leq 29.2$. More recently, \citet{2014ApJ...782....1S} and \citet{2014ApJ...792...60Z} used this method for fast moving NEOs, and \citet{2015AJ....150..125H} used it to search for MBAs. In the last year, authors have implemented faster searches with graphical processing units (GPUs) and massively parallel analyses capable of searching a finer grid and stacking a deeper stack of short exposures \citep{2019arXiv190711299Z, 2019AJ....157..119W}. Additionally, digital tracking has been used to estimate more precise astrometry, which is vital for fast moving objects for accurate orbit determination \citep{2018AJ....156...65Z}. Most recently, \citet{2019AJ....158..232H} has used it to conduct a complete search for MBAs down to $m_V=25$. 

In each of these studies, a linear assumption was made in order to avoid smearing the signal, which is an inherent cap on the usefulness of this technique. In this paper, we will describe a method that relaxes this requirement and enables a fully blind search for moving objects within survey data. A fully implemented version of our method could bring wide-area, long time-domain searches into play for digital tracking methods. We develop this method with an eye toward LSST \citep{2019ApJ...873..111I}, which will probe 2$\times10^5$ square degrees to a $\text{SNR}=5$ depth of $m(r)=27.5$ by stacking $\sim1000$ images per pointing in six filters (ugrizy). It will likely detect hundreds of thousands of asteroids over the ten-year survey with single-epoch detection methods. Our method is suitable for detecting asteroids in wide field surveys well below the single-epoch detection limit while also recovering valuable orbital and astrometric information simultaneously. In \S\ref{sec:data} we describe our test data from the Dark Energy Camera \citep[DECam;][]{2015AJ....150..150F} mounted on the 4m Blanco Telescope at Cerro-Tololo Inter-American Observatory (CTIO) in Chile. In \S \ref{sec:method} we describe our detection method. We analyze our test data and present our results in \S\ref{sec:results}. Finally, in \S\ref{sec:discussion} we offer a discussion of future extensions to this method and contextualize our simple test here in this larger picture. 

\section{Data} \label{sec:data}

\subsection{Observations}

\begin{figure}
    \centering
    \includegraphics[width=\columnwidth]{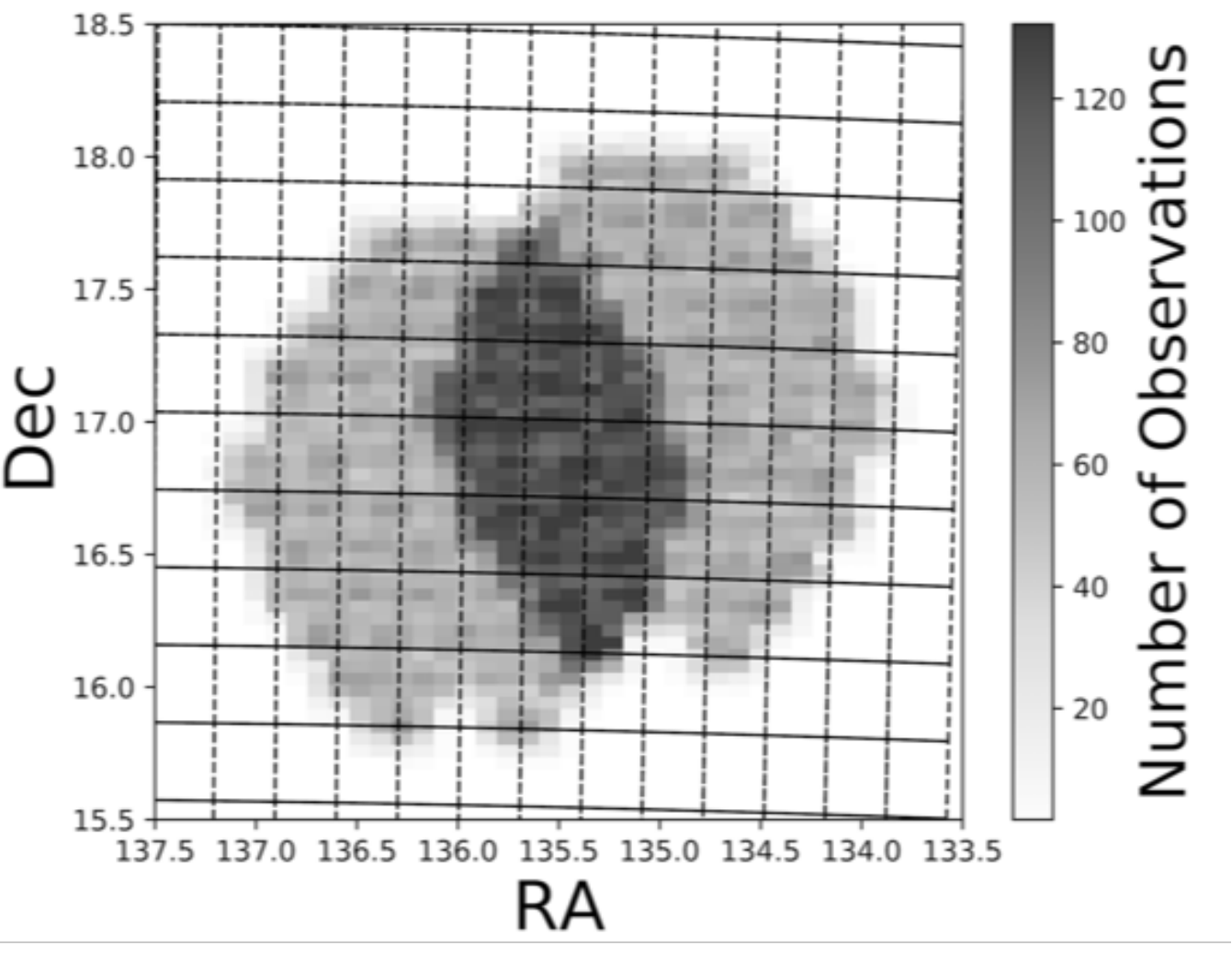}
    \caption{The data cube depth (number of frames) in R.A. and Dec. To generate the data cube, we cut the entire region into ``patches'' with the LSST data management pipeline, followed by creation of a master pixel grid. All analyses occurred on this grid.}
    \label{fig:observations}
\end{figure}

We obtained $\sim 3$ hours of observations on the Blanco Telescope at CTIO with DECam. These were spread over two consecutive nights (2018 December 22 and 23) during a 90 minute gap in the observability of the scheduled observations (Program 2018A-0273; PALS, PI William Dawson). The moon was full on 2018 December 21, so the nights were in bright time. In each night, we observed a single pointing toward the ecliptic plane, separated by a pro-grade translation of 1$\degr$ day$^{-1}$ along the ecliptic. Each night consisted of a $\sim 90$ minutes stack of 40 second exposures in a single filter. We dithered randomly by selecting pointings from $\mathcal{N} (\mu_{RA},\sigma_{RA}^2)$ and $\mathcal{N} (\mu_{DEC.},\sigma_{DEC}^2)$, where $\mu_{RA}$ and $\mu_{DEC}$ are  listed in Table \ref{tab:observations} and $\sigma_{RA} = \sigma_{DEC} = 2\arcmin$. The pointings over the two nights are shown also in Figure \ref{fig:observations}.

\begin{deluxetable*}{cccccc}
\tablenum{1}
\tablecaption{Summary of observations\label{tab:observations}}
\tablewidth{0pt}
\tablehead{
\colhead{Date} & \colhead{Filter} & \colhead{Exposures} & \colhead{$\mu_{RA}$} & \colhead{$\mu_{DEC}$} & \colhead{Median Seeing} \\ 
\colhead{\,} & \colhead{\,} & \colhead{Number} & \colhead{$^{\circ}$} & \colhead{$^{\circ}$} & \colhead{$\arcsec$}}
\decimalcolnumbers
\startdata
2018 Dec 22 & r & 85 & 135.00 & 17.04 & 0.97 \\
2018 Dec 23 & r & 82 & 136.01 &	16.77 & 0.87 \\
%2018 Dec 26 & VR & 88 & 138.99 & 15.88 & 0.93 \\
%2018 Dec 31 & VR & 88 & 143.92 & 14.33 & 0.87 \\
\enddata
\end{deluxetable*}

\subsection{Data Reduction}\label{subsec:reducution}

We obtained our exposures after standard processing by the DECam Community Pipeline \citep{2014ASPC..485..379V}. We then used the LSST software package\footnote{https://pipelines.lsst.io} \citep{2019ASPC..523..521B} to produce difference images. This can be split into three different pieces: making calibrated single epoch catalogs and images, creating a static template image, and subtracting the template from the single epoch images to produce difference images.

The LSST stack single epoch processing includes: the masking of cosmic rays, measuring the PSF, detecting objects, deblending and measuring individual sources and calibrating the astrometry and photometry. A detailed explanation of each of these steps is explained in the HSC data release papers \citep{2018PASJ...70S...8A,2019arXiv190512221A}. For calibration purposes we used the Pan-STARRS catalog \citep{2016arXiv161205243F}.

To create the template images we first interpolated the data onto a tangent plane projection centered on the average pointing. The total observed area was divided into a regular grid with each grid $4000\times4000$ pixels or $\sim 17\arcmin$ on a side (see Figure \ref{fig:observations}). For the templates we selected the best seeing images which were those with $\text{seeing} < 1.2\arcsec$.  For each grid the static sky is constructed following the procedure outlined in \citep{2018PASJ...70S...8A,2019arXiv190512221A}. Briefly,  the procedure builds a two-sigma clipped coadd to construct a static image of the sky. We subtract this coadd from each individual image and identify variable sources. Those detections that are truly variable will only appear in a small subset of the single epoch visits. The variable sources are then masked and a coadd image is created by taking the mean of all the images.

We use the Alard-Lupton algorithm \citep{1998ApJ...503..325A} as implemented in the LSST stack to create difference images. This procedure estimates a convolution kernel which, when convolved with the template, matches the PSF of the template with that of the science image by minimizing the mean squared difference between the template and science image. The Alard-Lupton procedure uses linear basis functions, with potentially spatially-varying linear coefficients, to model the matching kernel which can flexibly account for spatially-varying differences in PSFs between the two images. The algorithm has the advantage that it does not require direct measurement of the images’ PSFs. Instead it only needs to model the differential matching kernel in order to obtain an optimal subtraction. 

After examining the difference images, we found that there remained a large number of artifacts due to problems in the difference imaging algorithm.  A majority of these objects were bright stars, but even fainter stars caused problems for our analysis.  We tried to remove these issues by masking out known objects.  We selected objects from PAN-STARS with $m_r < 20$ and masked them using the procedure described in \cite{2018PASJ...70S...7C}.  For each star the mask was composed of a magnitude-dependent circle for the star and a rectangle for the bleed trail.  We visually tuned the size of the circle and rectangle to match our data.  We split the data into two class above and below 14th magnitude.  The size of the radii in arcseconds are: 
\begin{equation}
    {\rm radius} = 
        \begin{cases}
          400 \, e^{-m_{r}/3.8}, & \text{for}\ 14<m_{r}<20 \\
          400 \, e^{-m_{r}/4.1} , & \text{for}\ m_{r}<14. 
        \end{cases}
\end{equation}The length and width of the rectangles are 1.5$\times$0.15 times the circles diameter and for $14<m_r<20$ and 6$\times$0.3 times the diameter for $m_r < 14$.  The longer rectangles were necessary to remove some long bleed trails for bright stars.

\section{Methods}\label{sec:method}

A typical shift and stack detection strategy is carried out by sampling a fine grid of angular rates and stacking multiple frames along this proper motion to produce a shifted median stack image, which is then used to detect point sources that are moving at the matched rate of the shifted stack. This is the optimal search methodology when a linear assumption can be safely made regarding the motion of the objects of interest and the grid is fine enough to ensure all point sources are detected as such. This is indeed the best method for analyzing our data in hand. However, we seek a more generic search method. If we imagine instead that the data cover a wide area of sky and are observed over a large time-baseline (e.g., from an LSST-like survey), no longer can the flux from a given asteroid that intersects many images be efficiently co-added in a manner like shift-and-stack since the linear assumption is no longer valid. Since a given set of orbits are not parallel trajectories through the data volume, multiple objects are not detectable simultaneously. This fact is what limits shift-and-stack methods from being useful beyond the time where non-linear motion occurs, but it also means that a search that moves beyond this time scale must be of a fundamentally different design. In this section we build the idea of sampling individual trajectories through a survey and combining the requisite pixels to gather all of the expected flux for an individual trajectory. This method, while not optimal for our data set, still provides relatively complete results (see \S\ref{sec:results}) in a straight forward manner. Ultimately, we will extend the methodology and code infrastructure built here for larger and wider surveys. 

\subsection{Prior Generation} \label{subsec:priors}

Still thinking generically, we sample from a set of priors that parameterize the motion through the data. For example, we could randomly sample Keplerian orbital elements (allowing $e>1$) if we aimed to search for objects on hyperbolic orbits, or we could sample randomly from elliptical orbits with perihelia smaller than 1.3 AU to tailor our search to NEOs. The task is then to compute the intersection of a given survey with a randomly sampled trajectory and determine if the randomly selected set of pixels constitutes a detection. Note that this is not an easy task. The number of trajectories needed to fill a wide area and long time baseline survey is exceedingly large. However, the individual computations needed for each sampled trajectory are rather small.  

For our data at hand, we can sample linear trajectories. Each is defined by four parameters: proper motions and intercepts in on-sky angular coordinates, respectively. The proper motions are selected from a prior distribution while the intercepts are chosen randomly from a uniform distribution that spans beyond the extent of our data volume such that rays that enter the cube on the side are allowed. For a given trajectory, the position was selected as a random valid pixel within the data cube. This was done to ensure that at least one image was properly intersected by the trajectory. The proper motions were selected separately from two different probability distribution functions--one for motion parallel to the ecliptic and one for motion perpendicular to the ecliptic. After these were selected, they were converted to equatorial and then to pixel coordinates.

\begin{figure}
    \centering
    \includegraphics[width=\columnwidth]{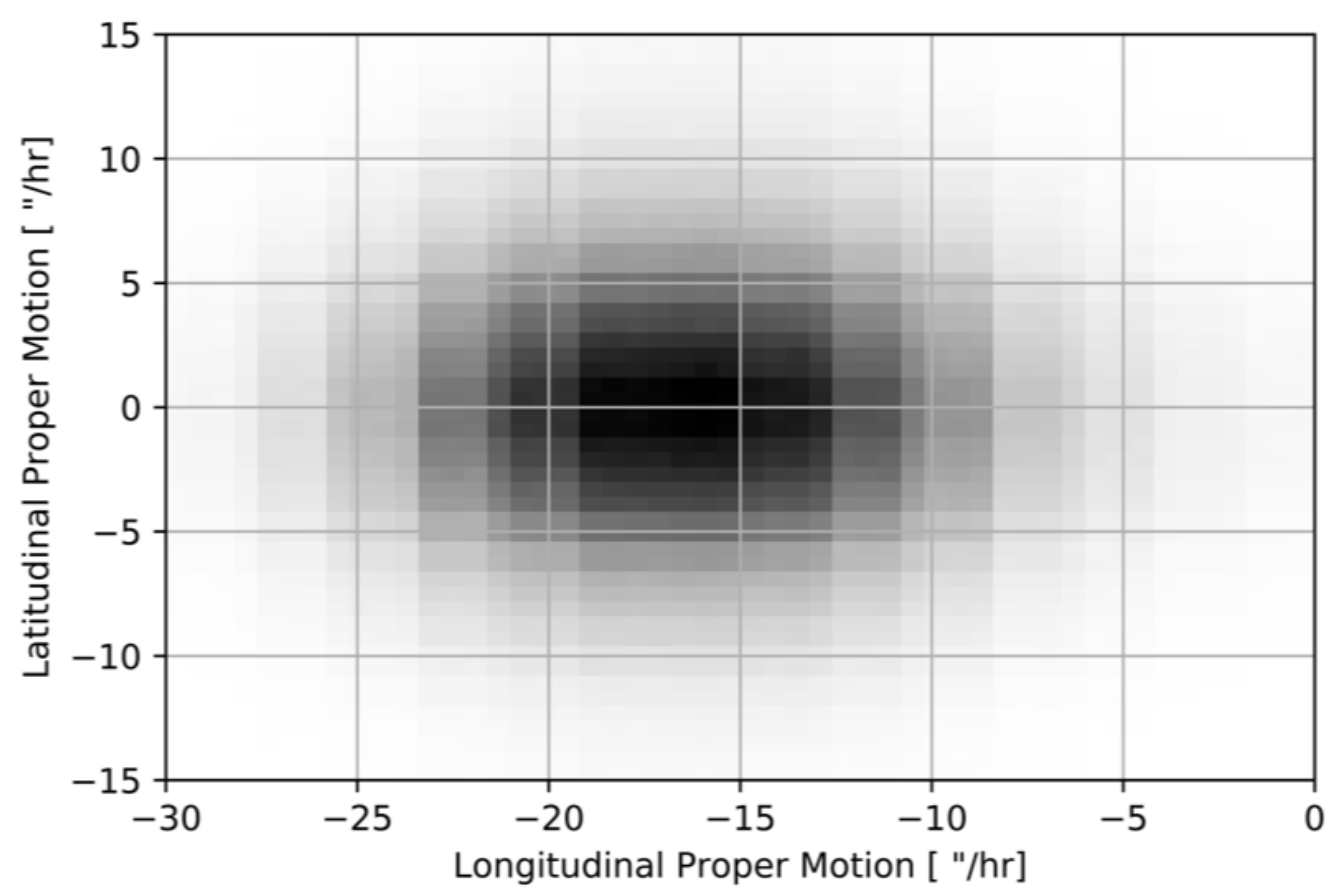}
    \caption{Proper motion distribution used for trajectory priors. This distribution was based on simulating all known objects in the MPC database observable in our survey. It was smoothed using a boxcar filter of width $5 \arcsec / hr$ and $9 \arcsec / hr$ in latitude and longitude respectively.}
    \label{fig:slope_distribution}
\end{figure}

We simulated both MPC and NEOs from \citet{2018Icar..312..181G} to generate prior distributions for the proper motions.  The simulated NEO population provided a more accurate distribution for NEOs since only a small, biased fraction of the MPC database contains NEOs. We randomly simulated orbits up our data space selecting those that were within the field of view (FOV) of DECam at the time of our observations. We chose to calculate proper motion priors in ecliptic coordinates because asteroid motion is predominantly along the ecliptic, and this allowed us to create the sharpest difference between the orthogonal components of the proper motion. The computed proper motion distributions are in Figure \ref{fig:slope_distribution}.

\subsection{Random trajectories to pixels}

In this subsection, we describe the manner in which random trajectories interact with the data in memory. This is a simple approach to efficiently accessing the relevant data in the case where it may all be loaded into random access memory (RAM). We will first describe the approach and then expand on possible extensions to it that could yield the requisite speed-up to make the method more applicable to larger data sets. 

We utilized the Flash cluster at Lawrence Livermore National Laboratory (LLNL) employing 2x Xeon E5-2670 v3 (12 cores each) CPUs and 256 GB of SSD memory. This is important to avoid having to access the data through the file system, which adds substantial latency to the queries. This also begins to approximate the requirements for larger survey searches, which will need to be distributed over many nodes of a large high-performance computing (HPC) system. This will add additional challenges such as node to node communication bandwidth and communication cross-linking, which we will discuss in more detail later in this subsection and in \S\ref{sec:discussion}.

In general, any time-domain survey data may be thought of as filling a three-dimensional space (two spatial and one time), where the units of the two-dimensional spatial domain are pixels covering the two-sphere and the time domain is discretized at irregular steps corresponding to the time of each exposure in the survey. In the linear regime this space may be simplified to a ``cube'' with square pixels and a short time axis. This assumption allows data collision and lookup indexing to be fast and relatively simple, but is inaccurate over wider surveys and longer periods of time -- this is the same challenge that makes traditional shift-and-stack fail beyond a short time regime. 

In our case, the survey data set is constructed into this conceptual cube by composing the data space as a three-dimensional stack of layers, with each layer itself being constructed from a series of ``tiles'' that come from the LSST data processing stack utilized in \S\ref{subsec:reducution}. Tiles contain the ``real'' surface of the data, i.e. the portions which exists somewhere on disk in a FITS file, and each FITS file in the data set is represented as a virtual tile in memory. The FITS files also contain information on the position and size in this global linear coordinate system and that information is used for positioning the corresponding tile within each layer.

The global linear coordinate space is partitioned at each level of the data cube hierarchy. At the top-level, the cube's two-dimensional projection is the minimum bounding box containing all of the tiles in the reduced data. The layers of the cube can be thought of as the intersection of the cube with parallel planes spaced at the time of each exposure (or perhaps thick planes with thickness corresponding to the exposure time). Each layer shares the same two-dimensional coordinate space (with small differences corresponding to dithering) and differ only in time. Layers also contain information about the length of the exposure represented by the tiles on that layer, so that the flatted two-dimensional footprint of a trajectory's intersection with the layer can be accurately calculated (this is important for using a proper signal-matched filter later). Tiles have a local coordinate space equal to that of the data array in the FITS file they represent.

Once a random trajectory is sampled, we propagate it through the data cube. At each time where an image exists in the cube, we calculate the trajectory--data cube intersection (in the linear assumption this is trivial), and we pull in the relevant pixel data for the SNR calculation. When looking up pixel data, each level in the cube hierarchy is queried to access the portion of the coordinate stored before moving to the remainder of the query to subordinate levels, if applicable, until the pixel values are fully resolved. All such queries are done by giving the cube object a point in three-dimensional space. First, we determine if the requested position is within the bounding box of the layer and if so, we then determine which layer was requested and pass the remaining two-dimensional query to that layer. We then determine  whether or not that point falls within a tile of the layer or within a chip gap or star mask. If the pixel queried exists and contains data, the query is passed to that tile. Once the query reaches the tile, the code retrieves the requested data from its associated FITS file. This is done by renormalizing the requested point to the local coordinate space of the tile and indexing into the data array segment of its associated FITS file to retrieve the requested pixel. A specially encoded NaN value is returned if the requested point cannot be further resolved to a real pixel at any point in this process (e.g., this is the case if the pixel lands in a chip gap or star mask).

In the interest of performance, we did not use any existing FITS library to perform the retrieval of pixel data. Instead, we use Linux's \textsf{mmap} (or memory map) feature to ``map'' the primary data array of the FITS file to a 1D array in memory. This provided a number of challenges to overcome but also proved significantly more performant than using, e.g., \textsf{CFITSIO}\footnote{\url{https://heasarc.gsfc.nasa.gov/fitsio/}} for this task. The main hurdle for our method was that FITS files store data records in (archaic) spans of 2880 bytes while the memory page size on a typical modern compute sysetm is 4096 bytes, so it is unlikely that the start of a FITS data record will be page-aligned on a modern computer. Linux's \textsf{mmap} only allows mappings that begin from file offsets that are multiples of the system page size. To work around this, we instead map the offset of the system memory page that contains the start of the data record we are interested in reading and note the difference between the two offsets when performing file accesses. FITS files may also store their data values in different forms of endianness, so we perform an endian swap when reading the pixel data if the endianness of the FITS file does not match that of the system.

\subsection{Number of trajectories}

Here we estimate the number of required trajectories to satisfactorily cover the space. Following \citet{2015AJ....150..125H}, the grid space needed to fully sample the proper motion parameters is given by:
\begin{equation} \label{eq:delta}
    \Delta = \frac{\sqrt{2} b_{max}}{t_{int}}, 
\end{equation}where $b_{max}$ is the maximum allowable blur ($\sim 1$ PSF) and $t_{int}$ is the total integration time of the stack (about 100 minutes in our case). Our prior has a radius of  $\sim 50 \arcsec$ s$^{-1}$ in two-dimensional proper motion (Figure \ref{fig:slope_distribution}), so a rough estimate of the required number of trajectories that must be sampled is given by the two-dimensional proper motion domain divided by the square of Eq. \ref{eq:delta} multiplied by the number of PSFs in a DECam FOV. This procedure gives $\sim 300\times10^{9}$. Computing detections for this number of rays took roughly six days on ten nodes of the Flash computing system. In \S\ref{sec:discussion} we will discuss improvements to our code that can vastly increase the sampling rate, even with larger data loads and more complex trajectories. 

\subsection{Signal-to-noise Ratio Calculations} 

Once the requisite pixels are pulled from the data, the SNR is calculated with a weighted sum using a two-dimensional Gaussian profile convolved with a line segment equal to the length of the expected streak, which we can compute from the sampled proper motion for each trial trajectory (note that for most of the trajectories this is very short, and the convolution reduces to the Gaussian PSF). The flux and noise in each pixel were summed across all images that a given source intersected. Within one image intersection, we measure the flux $F$ by calculating a weighted sum over the pixels $(p)$ inside the streak: 
\begin{equation}
    F = \sum_{i}{w_i * p_i}. 
\end{equation}The weights $(w)$ are determined from the profile, which is the convolution of a line segment with a two-dimensional Gaussian. The weights are normalized such that their sum equals the effective number of pixels in the streak: 
\begin{equation}
    n_{eff} = \sum_{i} w_i. 
\end{equation}The effective number of pixels is calculated by integrating the flux profile:
\begin{equation}
    n_{eff}^{-1} = \int{dxdy\,P(x,y)^2}. 
\end{equation}Specifically, we consider a streak with length $L$ oriented along the x-axis, width $\sigma_y$, flux per unit length $l_0$, and center (0,0), convolved with a symmetric bivariate Gaussian PSF of width $\sigma_{\pi}$: 
\begin{multline}
    P(x,y) = \frac{l_0}{2L} 
    	\frac{1}{ \sqrt{2\pi(\sigma_{\pi}^2 +\sigma_y^2)} } 
    	\exp\left(
    		\frac{-y^2/2}{\sigma_{\pi}^2+\sigma_y^2}
    	\right) \,
    	\times \\
    	\left(
    		\erf{\left(
    			\frac{L - 2x}{2\sqrt{2\sigma_{\pi}^2}}
    		\right)} + 
    		\erf{\left(
    			\frac{L + 2x}{2\sqrt{2\sigma_{\pi}^2}}
    		\right)}
    	\right).
\end{multline}
Assuming that the object is unresolved, $\sigma_{\pi} >> \sigma_{y}$, this simplifies accordingly:
\begin{multline}
    P(x,y) = \frac{l_0}{2L}\frac{1}{\sqrt{2\pi\sigma^2}}\exp{\left({\frac{-y^2}{2\sigma^2}}\right)}
\,\times \\\left(\erf{\left(\frac{L - 2x}{2\sqrt{2\sigma^2}}\right)} + \erf{\left(\frac{L + 2x}{2\sqrt{2\sigma^2}}\right)}\right),
\end{multline}where we have defined $\sigma \equiv \sigma_{\pi}$.

We choose an area surrounding the streak to sum over that balances capturing as much signal as possible with limiting computation time. We use twice the full-width-at-half-maximum of the point spread function ($2\sigma$) as a limiting radius away from the streak's ridge to integrate to, which encompasses over 99 percent of the signal.

Sources of noise include sky background Poisson noise, read noise, dark current, and Poisson noise from the signal itself. The total noise $N$ includes each of these but is dominated by the sources of Poisson noise, which add in quadrature. Thus we have:
\begin{equation}
    SNR = \frac{F}{\sqrt{F + N}}.
\end{equation}The background noise is calculated as the standard deviation of pixel values in the difference image. The expected value per pixel will then be the square of this number, and thus we get the total background signal will be:
\begin{equation}
    N = \sigma_{\rm pixel}^2\, n_{\rm eff}. 
\end{equation} All together we have:
\begin{equation}
    SNR = \frac{F}{\sqrt{F + n_{\rm eff}\,\sigma_{\rm pixel}^2}},
\end{equation}where
\begin{equation}
    F = \frac{\sum_{i}{(l_i \, p_i)} \, \sum_{i}{l_i}}{\sum_{i}{l_i^2}}, 
\end{equation} This is derived from equations 2, 3, and 4, where $l_i$ is the formula for the profile and the sums are taken over pixels, which is an approximation for the integral.

\subsection{Cuts and Outlier Rejection}

Before calculating the overall SNR, some initial filtering is required to deal with potential image artifacts intersected at a small number of exposures. To start, we store a pseudo ``light curve'' for each trajectory, which is the local SNR on each image the trajectory intersects. Even a few outliers in the light curve can dominate the overall computed SNR for faint asteroids, which are often well below single-epoch detection thresholds, so we apply some outlier rejection methods. We start with a rolling-median sigma clip, and then also apply an overall sigma clip (as opposed to a rolling-median) to trajectories that only intersect 20 or fewer images. We then calculate the total SNR for the remaining unclipped images. We require at least five images intersected and $SNR>5$. After this point, a trajectory is considered a candidate and we store its requisite information: proper motions and intercepts in pixel $(x,y)$ coordinates, overall SNR, and frames intersected. Many of these candidates are false positives or duplicates that are later filtered out.

Next, we apply a filter that removes duplicate detections of the same source. This is a frequent occurrence, especially for bright asteroids because a trajectory that only slightly intersects the deposited flux can quickly surpass the SNR threshold along only a portion of the true trajectory. This filter removes the vast majority of the candidate list. We use a k-d tree to find nearest neighbors for all the trajectories in the four dimensional space of $(x,y)$ position and proper motion (the proper motions are converted to spatial coordinates by multiplying by the total integration time of the data for each night). Using this four-distance metric, we consider all trajectories within $\epsilon=15\arcsec$ of other duplicates. To be explicit, we require:
\begin{equation}\label{eq:4dist}
    \sqrt{\left(|\vec{x}_{0,i}-\vec{x}_{0,j}|\right)^2 + \left(\Delta t \, |\vec{\pi}_i-\vec{\pi}_j|\right)^2} < \epsilon, 
\end{equation}where $\vec{x}_{0,ij}$ are the position of two trajectories at the first exposure, $\Delta t$ is the time between the last and first exposure, and $\vec{\pi}_{ij}$ are the proper motions. 

\begin{figure}
    \centering
    \includegraphics[width=\columnwidth]{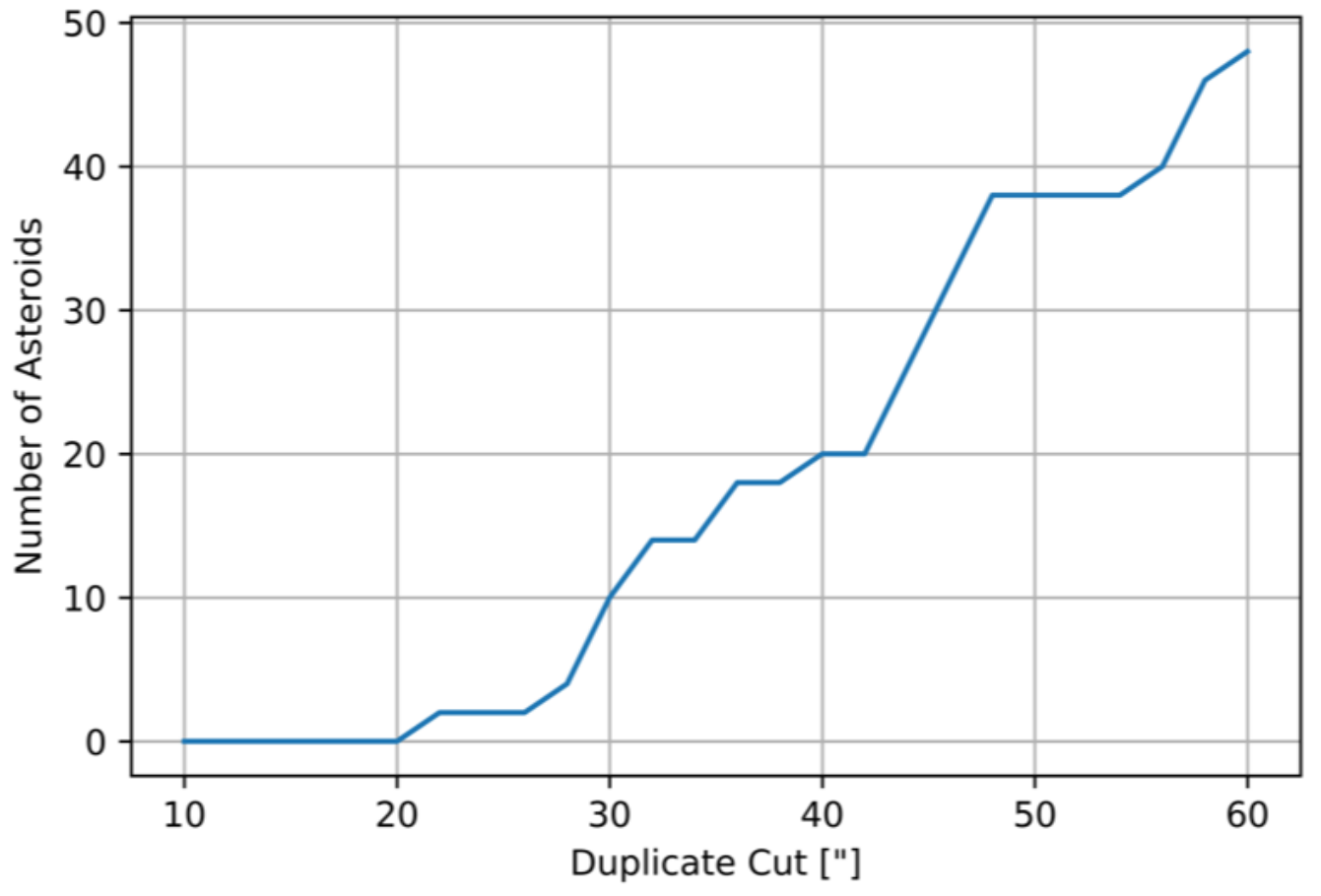}
    \caption{The number of asteroids removed from MPC catalog in our survey as a function of the size of the duplicate cut. No two asteroids are within 20$\arcsec$ in this catalog; there are likely some faint asteroids removed with a cut this large. We use a 15$\arcsec$ duplicate cut in the first step.}
    \label{fig:expected_asteroids_dist}
\end{figure}

We keep the trajectory with the highest SNR. We chose $\epsilon=15\arcsec$ as a balance between leaving in too many duplicates and cutting out trajectories that were not duplicates but simply close to each other; 15$\arcsec$ is small enough such that no two known asteroids from the MPC database that are expected to appear in our data are within this range of each other (see Figure \ref{fig:expected_asteroids_dist}). It is still likely that some duplicates pass this cut. We address these later.

We refine each trajectory to more properly align with the source using a Monte Carlo method. This is done by selecting a small postage stamp cutout around each trajectory's intersection with an image. We then sample $10^6$ trial trajectories within these postage stamps centered on the original detection trajectory (specifically, a bi-variate Gaussian of 3$\arcsec$.6 hr$^{-1}$ and 2$\arcsec$.6 width along the proper motion and position axes, respectively). We retain the highest SNR trajectory of the sample as the new, refined trajectory. We then apply a second duplicate filter setting $\epsilon=8\arcsec$. Any duplicates that passed the first cut will get closer to each other, and their source, though the refinement process. Thus, the second duplicate process can use a more stringent duplicate cut to successfully filter out almost all remaining duplicates.

Next, we apply a series of filters on the median stack image. This median image is calculated by taking the median across the postage stamp cutouts for each candidate. In this median image, a trajectory that perfectly aligns with a source would appear as a centered single PSF (or short streak for fast moving sources). First we calculate the SNR of this median stack using a weighted sum, and we require $SNR_{med}>7$. Next we apply two filters intended to cut out false positives that are a result of artifacts that survived the difference imaging procedure (typically near bright star masks or the edge of the DECam FOV). First we calculate the noise of the median stack image as the sigma clipped standard deviation of pixel values, and require that it is less than 10 ADU. Next we take all the pixel values outside the source profile and calculate the $\chi^2$: 
\begin{equation}
    \chi^2 = \frac{1}{N_{pix}}\sum{\frac{p_i^2}{N_i^2}}, 
\end{equation}and require it to be less than 1.5.

\begin{figure}
    \includegraphics[width=\columnwidth]{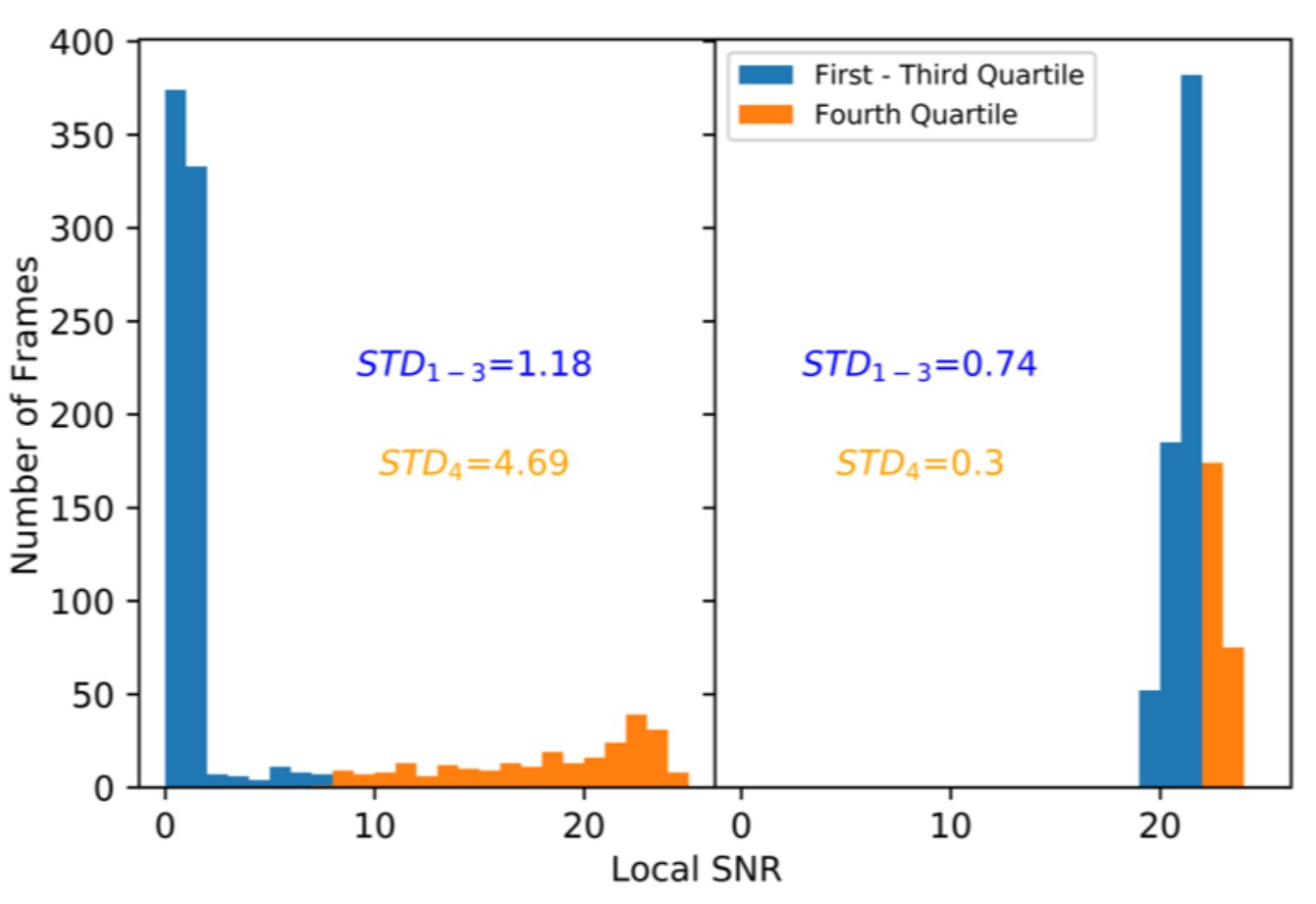}
    \caption{Example distributions of Local SNR values of frames hit for detection trajectories that are bad (left panel) and good (right panel) approximations of the true trajectory. The near hit filter compares the STD of the first through third quartile (blue bins) with the STD of the fourth quartile (orange bins). The near-hit mostly has local SNRs around 0 and some reaching higher corresponding to when the trail trajectory overlaps the flux from the asteroid, and so fails the test. The right panel shows a good hit, which has mostly non-zero local SNRs and passes.}
    \label{fig:nearhit_filter_example}
\end{figure}

Next, we apply a ``near-hit'' filter that attempts to cut out any trajectory that only partially overlaps with a real source in the data. These are a consequence of under sampling the trajectory priors, especially in the wings of the proper motion prior. We pause here to note that our primary goal is not to optimally detect asteroids in this trial field. We are more interested in developing a framework that may be properly scaled to larger data sets, and we are using this small test survey to being that development. We will return to this discussion in \S\ref{sec:discussion}. The near-hit filter is carried out by analyzing the distribution of local SNR values in the light curve. For a good hit, most of the local SNR values would distribute around a mean value due to Poisson noise in the flux and the sky background as well as intrinsic variability due to rotation of the asteroid. For a near-hit, the portion of the trajectory that misses the true asteroid motion would exhibit near-zero SNR. To differentiate between these two, we rank-order the local SNR values into an array and compared the standard deviation of the highest values of the distribution (fourth quartile) to the standard deviation of the lowest (first through third quartile). We required that the fourth quartile variance be less than the first through third quartile variance, thus requiring that most of the high value local SNRs are clumped around a value indicative of a detection on all of the frames that intersect the trajectory. We demonstrate asteroids that both pass and fail this cut in Figure \ref{fig:nearhit_filter_example}. In the case where all values are clumped around zero, and this filter is passed, the overall SNR would likely be below our threshold in the first place. 

Finally, we also apply a similar filter where we take each trajectory and randomly split its light curve in two. We then require that each half individually must reach the SNR cut of $\frac{5}{\sqrt{2}}$ to check that flux is rougly split between two random halves of the lightcurve. 

\begin{figure}
    \includegraphics[width=\columnwidth]{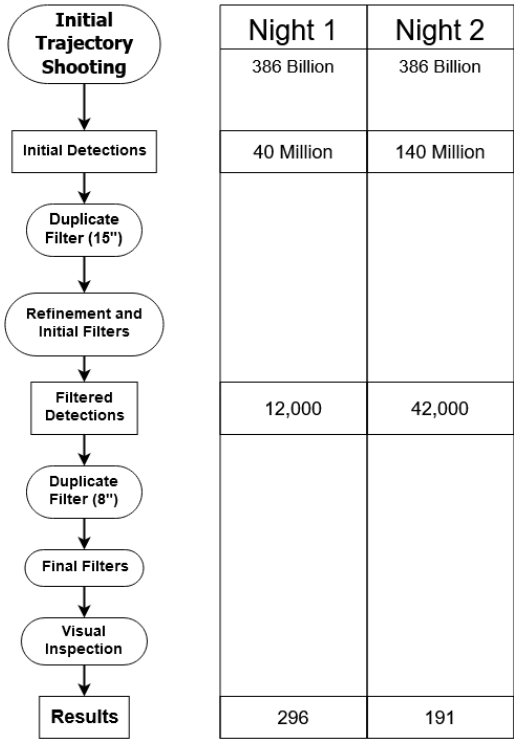}
    \caption{The flow chart for our filtering process along with numbers of detections for each night after every stage. The ``Initial Filters'' consist of just the median image filters, while the ``Final Filters'' consist of median image and light curve filters.}
    \label{fig:flowchart}
\end{figure}

We then visually inspect each result to remove any remaining false positives or potential duplicates. In figure \ref{fig:flowchart} we demonstrate the overall process as well as the number of detections after each stage for each night.

\section{Results} \label{sec:results}

\subsection{Injection and completeness}

During data reduction, we injected fake sources into the data to allow us to measure completeness. Roughly 1700 sources were injected each night with uniform distributions of proper motions up to 360$\arcsec$ $\text{hr}^{-1}$ and apparent r-band magnitudes between 18 and 28. We measured completeness by calculating the percentage of injected sources that were detected with respect to both magnitude and proper motion, where the latter is heavily affected by proper motion priors and refinement.

\begin{figure}
    \includegraphics[width=\columnwidth]{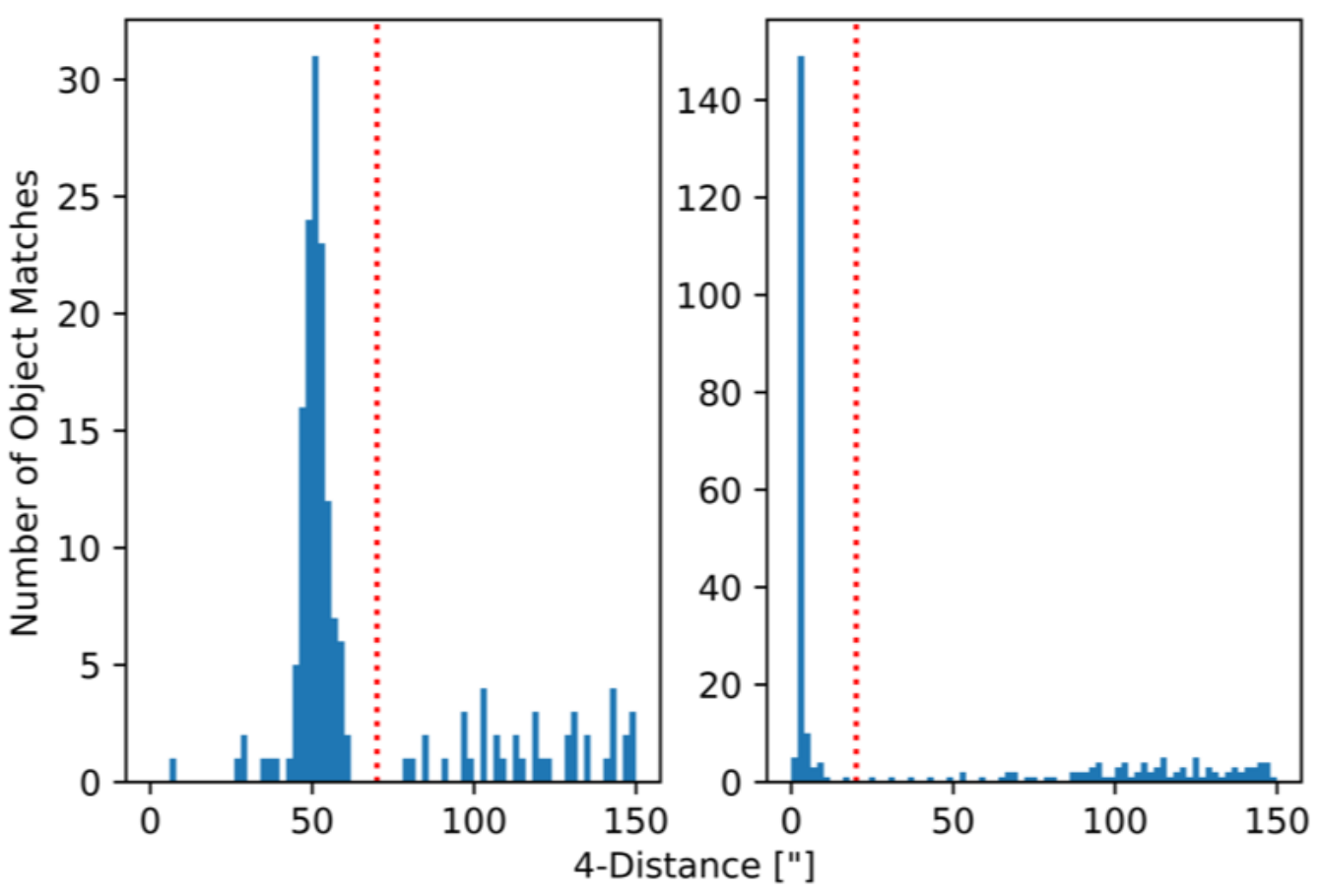}
    \caption{The distribution of four-distances from each detection to the nearest intended matching target (MPC objects on the left panel and injected sources on the right). There is a clear demarcation for each population; the red dotted line shows where we have chosen to separate matches from unmatched objects. These distributions are for night two only and has been cropped to show the small distances. Night one has similar distributions.}
    \label{fig:matching_distance_distribution}
\end{figure}

We determined which detections corresponded to injected sources using the same four-distance metric from Eq. \ref{eq:4dist}. For each detection, we found the closest injected source, and if it was within 20$\arcsec$, we considered it a match. We performed a similar analysis on known asteroids in the MPC database. We took all MPC asteroids expected to be within our observations and matched them with a larger minimum four-distance of $70\arcsec$. This was larger to accommodate errors in MPC orbital elements and potential orbit perturbations since last observation. We chose these two four-distance values empirically based on distributions of minimum four-distances, where there was a very clear demarcation between matched and un-matched sources (see Figure \ref{fig:matching_distance_distribution}). 

\begin{figure}
    \includegraphics[width=\columnwidth]{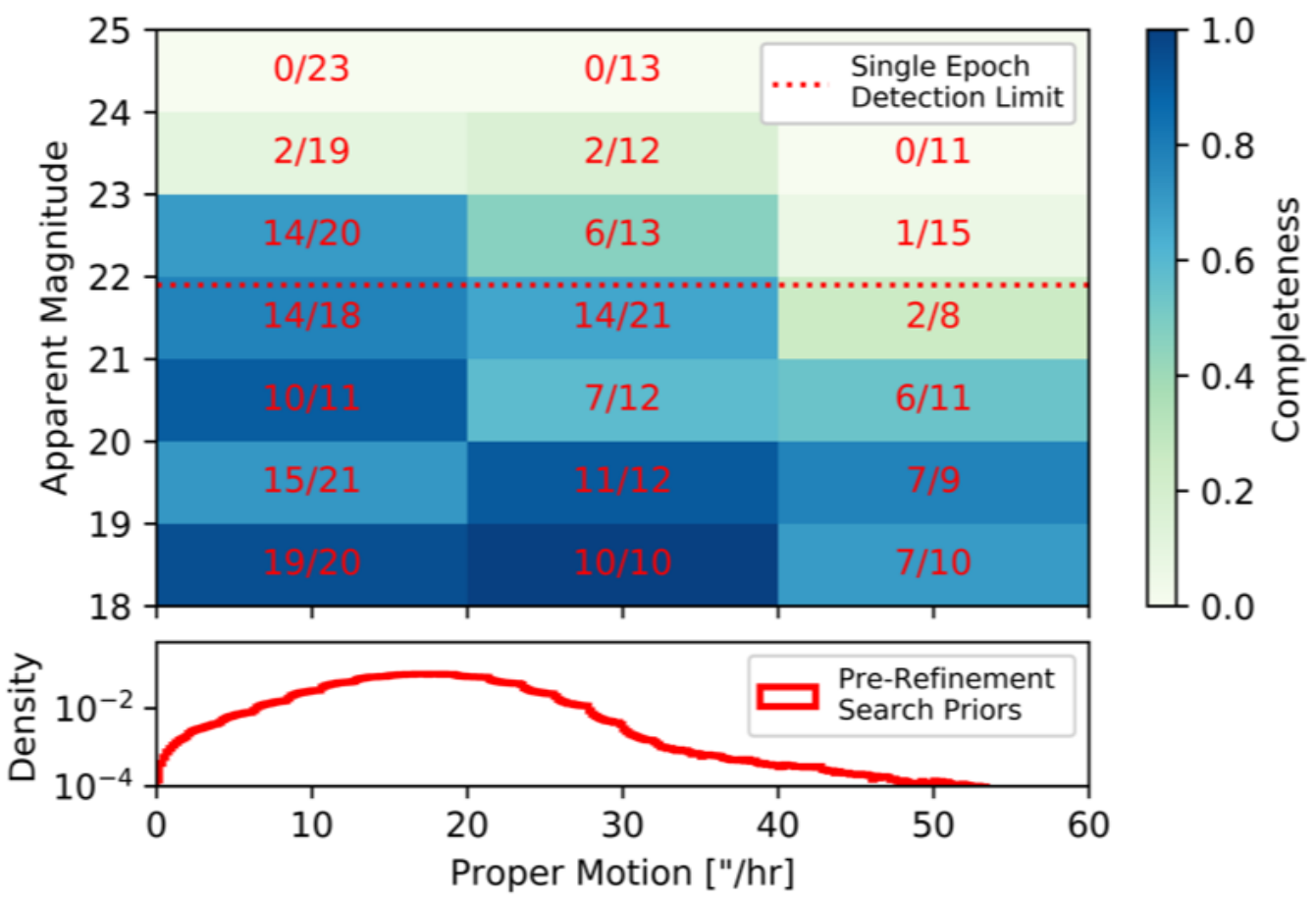}
    \caption{Night one injected source recovery. The red dotted line represents the 5 sigma limiting magnitude for single epoch detection (the limit drops minimally over the x-axis shown). The proper motion priors shown below heavily influence completeness with respect to proper motion. The detection of fast moving objects is a result of imperfect initial detection with possible adjustment with refinement.}
    \label{fig:completeness1}
\end{figure}

\begin{figure}
    \includegraphics[width=\columnwidth]{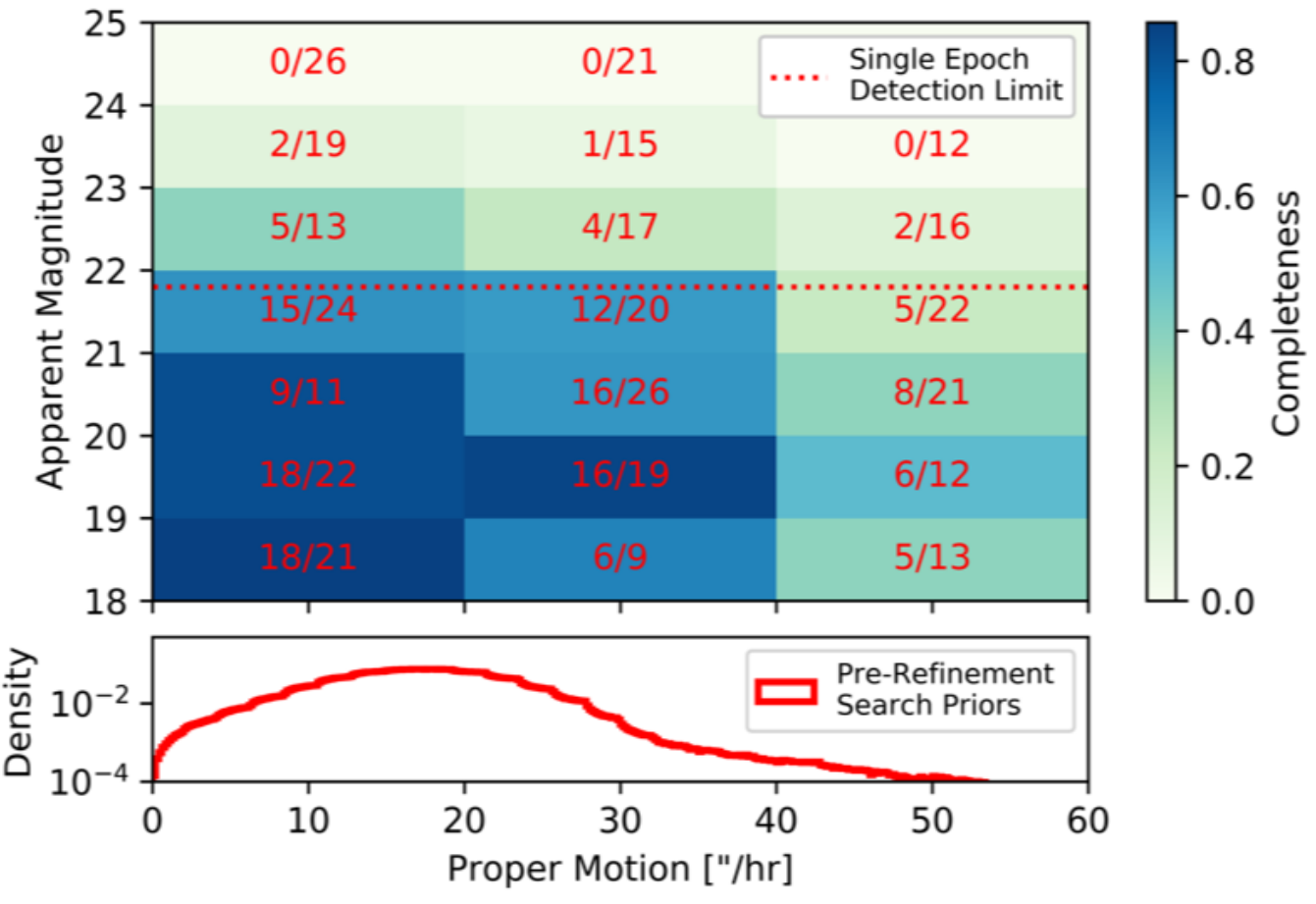}
    \caption{Same as Figure \ref{fig:completeness1} but for night two.}
    \label{fig:completeness2}
\end{figure}

In figures \ref{fig:completeness1} and \ref{fig:completeness2}, we show completeness along with our proper motion priors. The 5 sigma, single-epoch detection limit was $m_{r}<21.9$ and $m_{r}<21.8$ for nights one and two, respectively. We determined the 5 sigma cuts using the DECam exposure time calculator with a sky background estimated using the model described in \citet{1991PASP..103.1033K}, which is derived based on Mie scattering of aerosols and Rayleigh scattering. The Mie scattering term is an empirical fit at large angles though \citet{1991PASP..103..645S} suggests it is valid to within roughly 10$^{\circ}$ of the moon, which is closer than our fields for each night. The lack of completeness below the red dashed line is suggestive of an incomplete search of the priors, which we mentioned was evident from the large number of near-hits. We again refer readers to \S\ref{sec:discussion} on this point, as our main goal of this analysis is to work toward an analysis framework that is scaleable to larger data sets on HPC resources. Indeed, the detections above the red dashed lines in Figures \ref{fig:completeness1} and \ref{fig:completeness2} suggest that the method is working. 

\begin{figure*}
    \includegraphics[width=\textwidth]{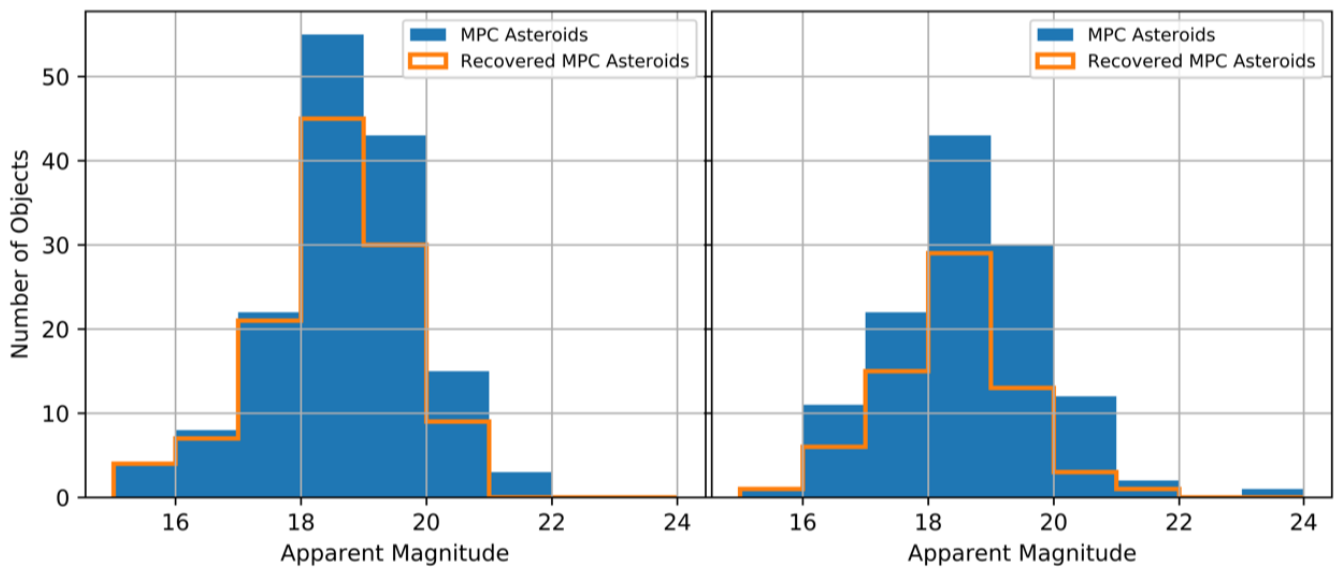}
    \caption{\emph{Left:} night one recovery of known MPC asteroids. The magnitude distribution of known MPC asteroids in the FOV is shown in solid blue. Those that were successfully recovered are shown in orange. We matched our detections with the known MPC asteroids using the four-distance metric on MPC asteroid positions propagated to the relevant epoch. \emph{Right:} Same for night two, as with the larger number of false positives, we find that night two was also less sensitive to MPC known asteroids.}
    \label{fig:completeness_mpc1}
\end{figure*}

In Figure \ref{fig:completeness_mpc1}, we show completeness specifically for known MPC asteroids. We successfully detect nearly all bright known asteroids and the majority of MPC objects brighter than $m_{V}=21$. These figures display only the detections that are matched with known MPC asteroids, while a substantial fraction of our detections are new.

\subsection{Detections}

\begin{figure*}
\centering
    \includegraphics[width=\textwidth]{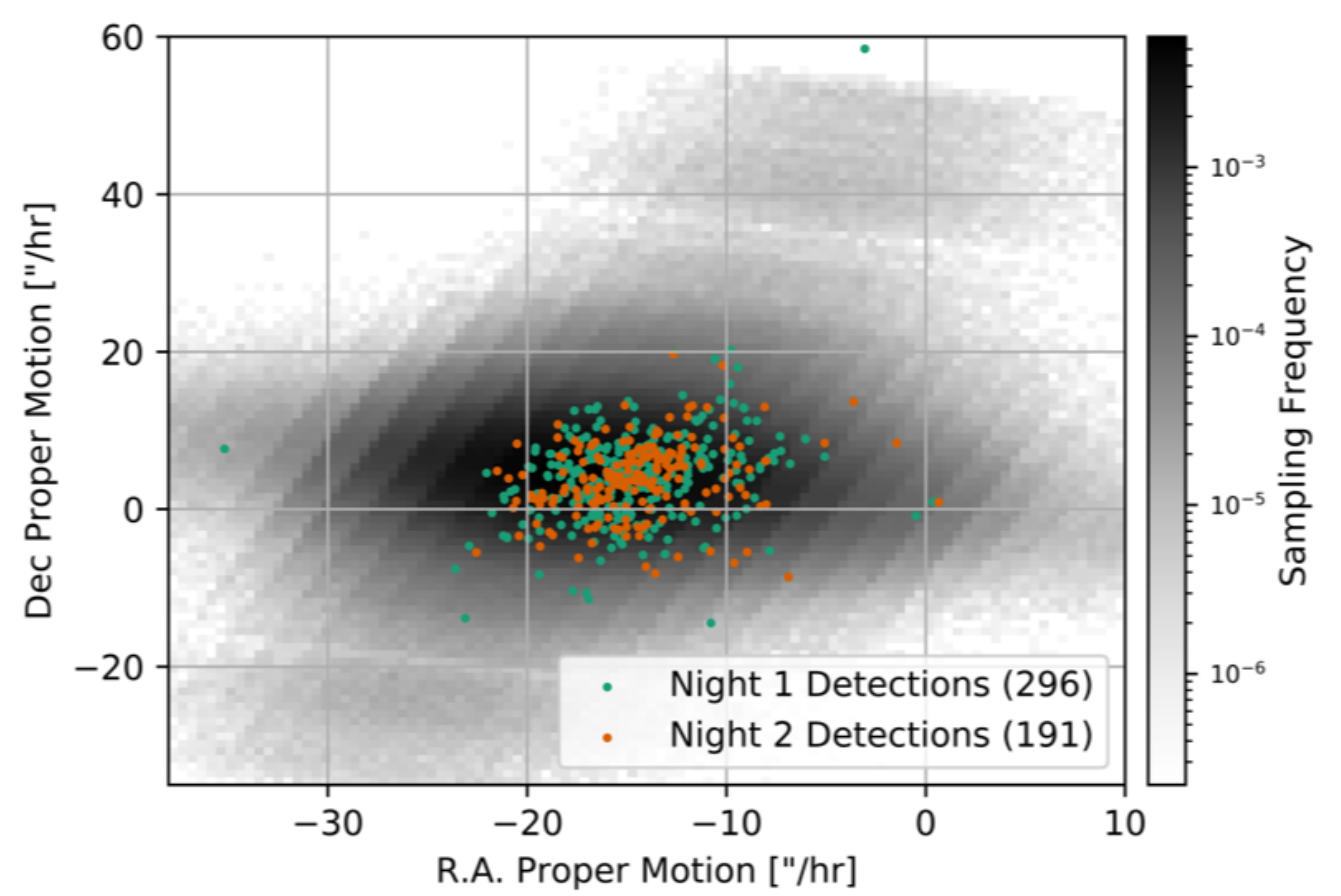}
    \caption{Proper motion of detected asteroids for both nights. The grey scale background shows the proper motion priors used in sampling trajectories. The occasional alignment of dots from each night is the result of detecting the same object across both nights.}
    \label{fig:detections}
\end{figure*}

\begin{figure*}
    \centering
    \includegraphics[width=\textwidth]{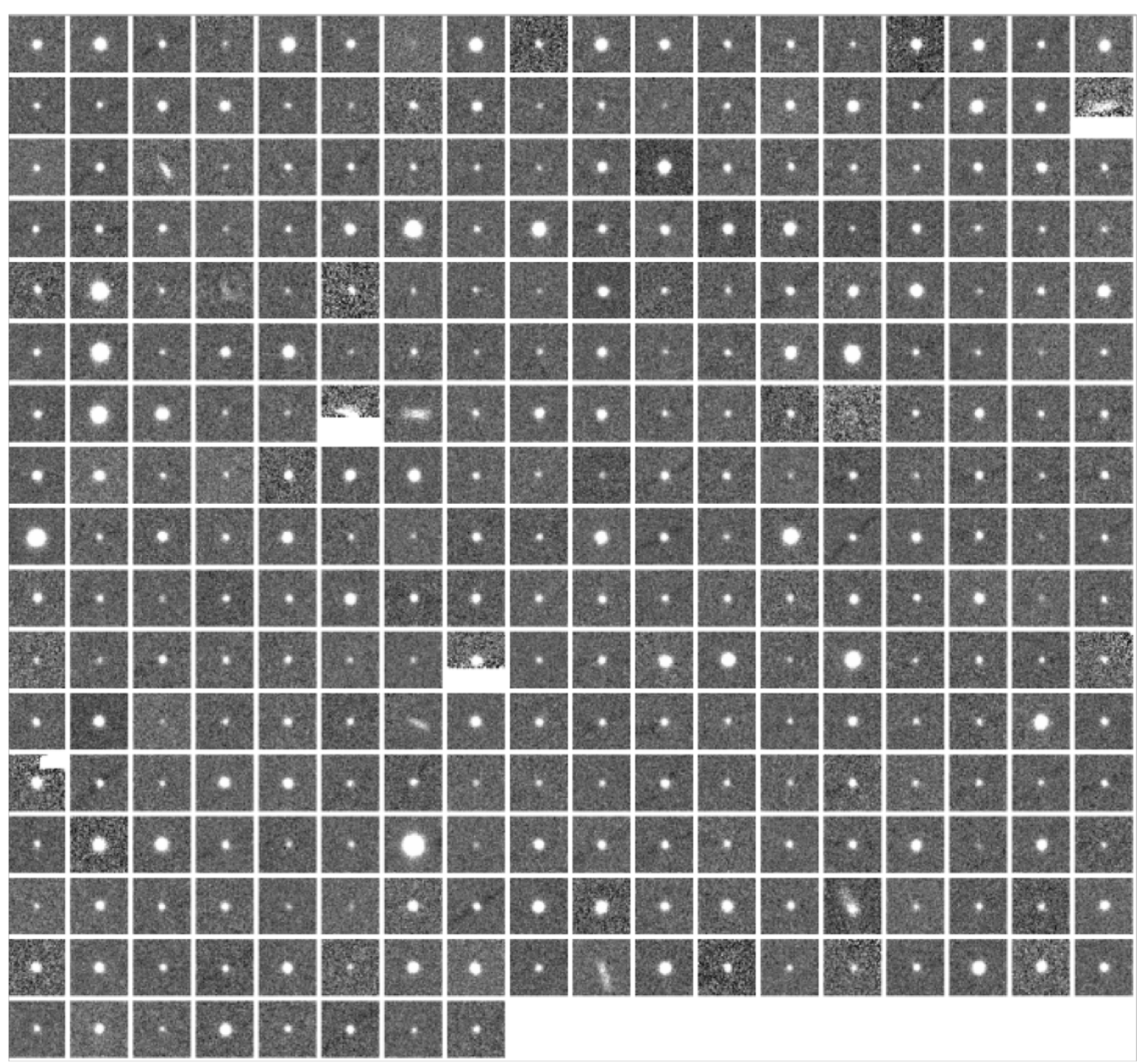}
    \caption{night one detections.}
    \label{fig:night1}
\end{figure*}

\begin{figure*}
    \centering
    \includegraphics[width=\textwidth]{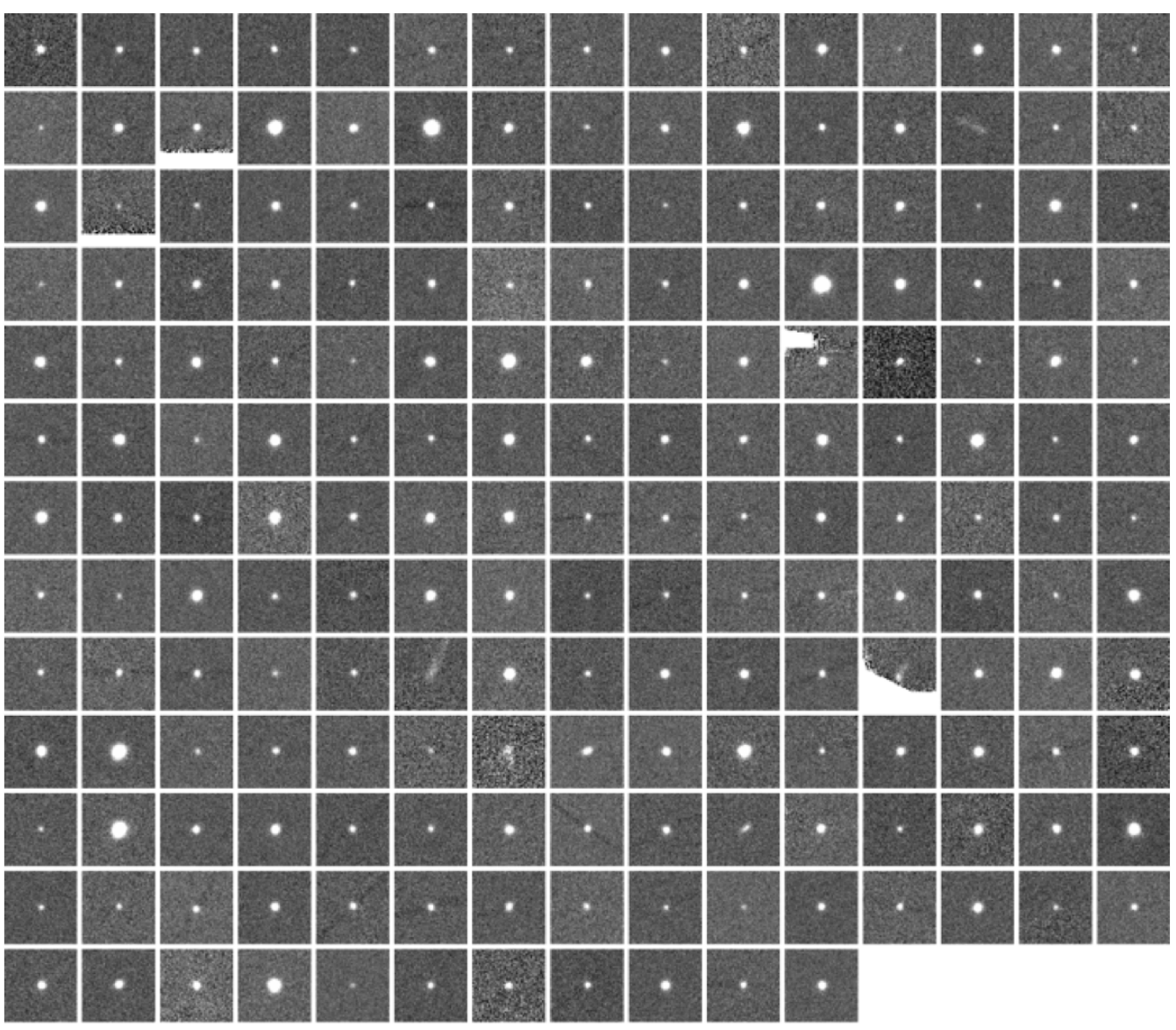}
    \caption{night two detections.}
    \label{fig:night2}
\end{figure*}

We detected a total of 487 asteroids across both nights, of which 301 are recovered MPC objects. The detections for each night are shown in Figure \ref{fig:detections} overlaid on the proper motion prior in log-space to emphasize the tails of the distribution. Two objects detected only in night one where detected in the tails of this prior. A thumbnail of the median stack along the detection trajectory is also presented for each detection for nights one and two in Figures \ref{fig:night1} and \ref{fig:night2}, respectively. 

In total, we detected 64 asteroids on both night one and night two. We selected these 64 asteroids by comparing detections across nights and enforcing the same flux was measured and the two detections occurred within $50\arcsec$ of the extrapolated position from the first night detections to the second night. All detections have been reported to the MPC. We present them in Tables \ref{tab:night1} and \ref{tab:night2}. We explored using the rotational reflex velocity (RRV) method to determine these asteroids distances \citep{2015AJ....150..124H, 2016AJ....152..183L}; however, because our observations were not taken near opposition, this method does not provide a one-to-one mapping to distance. 

\section{Discussion} \label{sec:discussion}

\subsection{HPC Considerations}
We have alluded to future use cases and extensions for our constructed pipeline throughout this paper, so in the spirit of full specificity, we will discuss the reasons for adopting such a pipeline, the benefits of it, and also the challenges that are unique to this framework.

Between several surveys (past, present, and future) the vast majority of the sky will be observed repeatedly over several decades. Regardless of flux, photons from a large fraction of all asteroids in the Solar System have been deposited in these exposures on hundreds if not thousands of occasions -- the vast majority of which may be too faint to detect in single-epoch detection schemes, thus making the embedded information useless for most asteroid detection and linking methods. Accessing this information and combining it to detect more asteroids in a given survey is our fundamental goal. In \S\ref{sec:method}, we laid out the basic idea behind the method. The key features of our method are sampling random trajectories under a prior distribution followed by a finely tuned mapping and gather operation on the data space. The gather operation quickly becomes the bottleneck because a single orbit touches numerous files, which must be accessed separately before the simple requisite computation (e.g., PSF convolution, weighted sum, etc.) can be made. We note here that this challenge is an impediment that traditional shift-and-stack methods do not suffer from because the underlying assumption of linearity ensures that parallel trajectories may always be tested simultaneously via detection on the shifted median stack. Here, we have relaxed the linear assumption and address the I/O challenge head on.

In this paper, to get around this challenge, we used a compute system that was capable of loading all of the requisite data into RAM. We thus avoided latency in accessing the files repeatedly over a file server. In the case where the data load is much larger, this will no longer be possible. A simple \textsf{mmap} implementation becomes much less efficient in the manner we carried out. We are now developing a batched method, which accepts the latency of multiple reads from the file server and instead spends RAM on simultaneous propagation of a larger number of trajectories through time. We have found comparable trajectory sampling rates per node but an ability to scale to arbitrarily large numbers of compute nodes. A follow-on paper will analyze the new implementation performance on a similar data set before we graduate to sampling non-linear trajectories and larger data volumes over longer time domains (Golovich et al. in preparation).

Unfortunately, many of the newest HPC clusters (and more scheduled in the future) do not enable the communication bandwidth to make such a search tractable on large surveys. We must address these issues in order to follow our approach to larger data volumes. As the data volumes grow, the separation of requisite pixels for a given SNR computation grows as well. That is, pixels distributed around an HPC system's RAM are needed to be combined. In HPC parlance, our code has poor \emph{locality}. Furthermore, since nearly any number of distinct trajectories touch each pixel, there is a high data access to computation ratio. This tends to add latency in a traditional HPC computation for data fetches. \citet{Lumsdaine06challengesin} describe these types of computational issues for the case of graph algorithms on large data sets, which share many challenges with our vision of a blind asteroid detection pipeline.

Longer time domains also present a new challenge. That is, the search space grows rapidly: a fractional change in orbital elements results in a larger projected, on-sky spatial separation over time. In fact, for large surveys (e.g., Pan-STARRS, ZTF, LSST), the number of possible unique trajectories through the data space becomes exceedingly large. 

Despite this search volume problem, the potential to detect extremely faint asteroids remains, so we now explore some of the requirements and challenges posed. Given the immense search space, the most likely use case for a code that can extract detections from a whole survey is in densely searching niche orbital regimes of particular interest such as Earth Trojan asteroids \citep[e.g.,][]{2011Natur.475..481C} or interstellar objects \citep[e.g.,][although we should note that the search space for these is even more immense given the unbounded eccentricity]{2017Natur.552..378M}. Under a narrow set of priors for the motion of specific types of objects, an efficient sampling, mapping, and gathering of relevant pixel data from an entire survey could lead to new discoveries. Expanding the discovery space requires optimization of each step (sampling, mapping physical trajectories to the pixel data, and gathering that data). 
We have shown in this study the importance of optimized loading of data into RAM. In larger surveys, loading all the data into RAM becomes impossible on smaller compute systems, necessitating parallel distributed memory methods, which are foreign to most astronomers. Cutting-edge compute systems have upwards of several petabytes of aggregate RAM, enough to hold even very large surveys; however, accessing this data across many hundreds of compute nodes brings about additional challenges. For one, because an individual orbit may touch hundreds of images across the survey yet those images are distributed through the RAM, there is very little localization. In order to efficiently access the small number of pixels needed to compute the SNR, e.g., an immense amount of cross-talk between compute nodes querying other nodes for their data. These issues will be explored in our future studies. 

%In order to explore the potential efficiency of such a communication scheme, we performed the following test. A typical asteroid touches tens to hundreds of frames and deposits flux on tens of pixels. We generated a few thousand random 4k by 4k arrays and simulated the propagation of trajectories through the intersection of a random subset of those frames. We then grabbed a random set of pixels from the intersecting frames and communicated the requisite pixel quantities that would be required for an SNR computation for each trial. We completed both a hard and soft scaling test on this algorithm on the Catalyst HPC system\footnote{\url{https://computing.llnl.gov/computers/catalyst}} at Livermore Computing at Lawrence Livermore National Laboratory. We distributed the simulated frames over the compute nodes involved and utilized a \textcolor{red}{\bf (Roger to write this part)} test that exhibited \textcolor{red}{\bf (results)}. 

\subsection{Summary}
In this paper we motivated and developed a prototype for a new type of blind search for asteroids in astronomical survey data. We provided a proof of concept analysis on a small three hour observation with DECam detecting blindly numerous asteroids fainter than the single epoch depth would permit. We achieve this by parametrizing the motion of asteroids of interest and randomly sampling from prior probability density functions for the motion model and mapping those trajectories to the data space and computing the optimal signal-matched filter for detection. We present the results of our detection pipeline on our test data, finding 487 asteroids with 301 of these matching to known asteroids in the MPC database and 186 being new discoveries. These detections were made despite challenging observing conditions made within $20\degr$ of a nearly full moon. We thoroughly discuss the prospects for such a code in larger surveys and explore the HPC requirements for present and future asteroid surveys. 

\startlongtable
\begin{deluxetable*}{ccccccc}
\tablenum{2}
\tablecaption{Night One Detections \label{tab:night1}}
\tablewidth{0pt}
\tablehead{
\colhead{MPC Designation} & \colhead{$m_r$} & \colhead{R.A.} & \colhead{Dec.} & \colhead{$\pi_{\alpha}\cos{\delta}$\tablenotemark{a}} & \colhead{$\pi_{\delta}$} & \colhead{Observation Time \tablenotemark{b}}  \\ 
\colhead{\,} & \colhead{Magnitude} & \colhead{$^\circ$} & \colhead{$^{\circ}$} & \colhead{$\arcsec \, hr^{-1}$} & \colhead{$\arcsec \, hr^{-1}$} & \colhead{MJD}}
\decimalcolnumbers
\startdata
 & 20.21 & 134.79317 & 17.34582 & 57.431 & 16.524 & 58475.300347 \\
5512 & 14.48 & 134.02066 & 17.10905 & 49.997 & 4.529 & 58475.300347 \\
 & 20.75 & 134.29134 & 17.63576 & 56.773 & -8.453 & 58475.301169 \\
326927 & 18.31 & 135.53799 & 17.57403 & 41.795 & -5.965 & 58475.301169 \\
422173 & 18.92 & 134.0232 & 17.14441 & 65.643 & 39.024 & 58475.300347 \\
377387 & 19.2 & 134.71131 & 16.80628 & 58.3 & 28.746 & 58475.300347 \\
 & 19.77 & 134.27197 & 16.42143 & 56.311 & -4.806 & 58475.301169 \\
 & 19.61 & 135.2303 & 17.9343 & 35.45 & 60.322 & 58475.300347 \\
 & 20.19 & 135.09889 & 16.94402 & 66.378 & -7.589 & 58475.300347 \\
 & 19.05 & 135.34779 & 16.84297 & 34.476 & -3.1 & 58475.300347 \\
 & 18.0 & 134.62291 & 16.47335 & 11.164 & 221.666 & 58475.300347 \\
501015 & 21.06 & 135.26071 & 17.70396 & 66.039 & 16.29 & 58475.300347 \\
100878 & 17.99 & 135.14364 & 16.62999 & 38.803 & 26.496 & 58475.300347 \\
 & 20.33 & 134.82584 & 16.46962 & 77.055 & 6.383 & 58475.300347 \\
 & 20.36 & 134.67973 & 17.80584 & 48.188 & 26.118 & 58475.300347 \\
229604 & 18.85 & 135.74152 & 17.25404 & 55.572 & 1.447 & 58475.300347 \\
 & 19.51 & 135.45125 & 16.83472 & 31.218 & 5.512 & 58475.301169 \\
369867 & 19.02 & 135.46566 & 16.95659 & 64.4 & 33.97 & 58475.300347 \\
501117 & 20.73 & 134.42677 & 16.95386 & 68.706 & 15.689 & 58475.300347 \\
 & 20.76 & 135.37694 & 17.89672 & 48.122 & -6.48 & 58475.301169 \\
242209 & 18.01 & 135.12496 & 16.23311 & 75.061 & 9.547 & 58475.301169 \\
497263 & 19.31 & 134.78628 & 17.06208 & 30.148 & 20.783 & 58475.300347 \\
42019 & 17.32 & 135.57697 & 17.02326 & 55.307 & 17.87 & 58475.300347 \\
125679 & 17.26 & 134.48391 & 17.87131 & 61.07 & 48.136 & 58475.300347 \\
563034 & 18.83 & 135.41812 & 17.29607 & 52.998 & 49.363 & 58475.300347 \\
62327 & 17.86 & 135.19413 & 17.99334 & 55.54 & 27.205 & 58475.300347 \\
783819 & 19.49 & 135.28325 & 16.90288 & 50.448 & -13.525 & 58475.300347 \\
99083 & 18.1 & 135.94501 & 16.56385 & 55.344 & -16.92 & 58475.301169 \\
559253 & 20.09 & 135.72067 & 16.58669 & 76.843 & 4.673 & 58475.301169 \\
315998 & 19.73 & 135.56097 & 16.99877 & 43.285 & 21.197 & 58475.300347 \\
636698 & 19.84 & 133.87031 & 17.0173 & 50.014 & 37.303 & 58475.301968 \\
180367 & 18.64 & 135.75742 & 16.67427 & 55.399 & 30.834 & 58475.300347 \\
199168 & 18.51 & 135.82092 & 17.68091 & 64.178 & -12.744 & 58475.300347 \\
287521 & 18.28 & 135.61687 & 17.63634 & 40.531 & 45.061 & 58475.300347 \\
320504 & 18.58 & 134.23352 & 16.64174 & 46.233 & -3.2 & 58475.300347 \\
545217 & 19.87 & 134.74767 & 17.43255 & 70.063 & -10.462 & 58475.300347 \\
166030 & 18.21 & 135.72696 & 17.19883 & 43.37 & 21.078 & 58475.300347 \\
496976 & 19.68 & 134.77711 & 17.69075 & 33.327 & 23.594 & 58475.300347 \\
110244 & 18.02 & 134.92272 & 17.62696 & 57.894 & 23.569 & 58475.300347 \\
575004 & 20.22 & 135.1702 & 16.49363 & 68.63 & 26.687 & 58475.301968 \\
214124 & 17.69 & 135.79507 & 16.93669 & 30.802 & 42.509 & 58475.300347 \\
350748 & 17.89 & 135.40597 & 17.87755 & 61.462 & -40.19 & 58475.300347 \\
 & 19.65 & 135.70601 & 17.40023 & 48.279 & 13.637 & 58475.301968 \\
307572 & 19.43 & 135.39642 & 17.39503 & 44.701 & 20.88 & 58475.300347 \\
77485 & 17.42 & 135.61702 & 16.26229 & 44.789 & 23.292 & 58475.301169 \\
321863 & 18.96 & 135.68024 & 17.80152 & 64.584 & 35.323 & 58475.300347 \\
 & 20.09 & 135.75291 & 16.95639 & 70.368 & -31.471 & 58475.300347 \\
 & 20.14 & 135.79085 & 16.79599 & 49.26 & 19.26 & 58475.300347 \\
 & 20.29 & 135.17341 & 17.07839 & 66.86 & 5.108 & 58475.301169 \\
507022 & 19.66 & 134.43732 & 17.21644 & 65.624 & 7.78 & 58475.300347 \\
48143 & 17.07 & 135.02526 & 17.83662 & 34.815 & 51.106 & 58475.300347 \\
102405 & 18.16 & 134.77825 & 18.03703 & 54.663 & 31.115 & 58475.300347 \\
162069 & 18.79 & 134.68747 & 16.88823 & 71.141 & 27.857 & 58475.300347 \\
 & 19.96 & 134.27535 & 16.95371 & 56.777 & 20.66 & 58475.300347 \\
506183 & 20.74 & 134.61118 & 17.42175 & 65.341 & 19.609 & 58475.300347 \\
254574 & 18.55 & 135.79797 & 17.14525 & 40.31 & 24.242 & 58475.300347 \\
107655 & 20.34 & 134.59861 & 16.16852 & 56.539 & -1.181 & 58475.300347 \\
173983 & 17.54 & 134.91364 & 17.69367 & 34.551 & 36.223 & 58475.300347 \\
655789 & 19.69 & 135.4815 & 16.84174 & 59.223 & 42.707 & 58475.300347 \\
367784 & 18.99 & 134.78362 & 17.65727 & 59.545 & 49.518 & 58475.300347 \\
724316 & 20.47 & 134.76377 & 16.83589 & 45.08 & 7.517 & 58475.300347 \\
190276 & 19.56 & 135.78347 & 17.03281 & 68.821 & 16.402 & 58475.300347 \\
522895 & 19.31 & 135.8193 & 17.59754 & 40.039 & 23.076 & 58475.300347 \\
745999 & 18.38 & 134.47392 & 17.58256 & 47.03 & 8.892 & 58475.301169 \\
 & 20.63 & 135.39451 & 16.85289 & 62.923 & 4.46 & 58475.300347 \\
474046 & 19.7 & 134.59682 & 17.4886 & 56.689 & 15.055 & 58475.300347 \\
 & 21.06 & 134.59864 & 17.24698 & 36.805 & 15.124 & 58475.301169 \\
429278 & 19.87 & 135.71753 & 17.18038 & 48.23 & 28.087 & 58475.300347 \\
170608 & 18.78 & 135.31984 & 17.33802 & 53.01 & 14.767 & 58475.302766 \\
61818 & 17.66 & 134.93991 & 17.67428 & 59.961 & 48.089 & 58475.300347 \\
 & 19.52 & 134.74307 & 16.2332 & 41.346 & 43.186 & 58475.300347 \\
214666 & 17.1 & 135.56074 & 16.50028 & 34.317 & 68.267 & 58475.300347 \\
125851 & 18.4 & 134.8533 & 16.95914 & 54.672 & 35.107 & 58475.300347 \\
 & 19.45 & 134.9119 & 17.41068 & 48.654 & 26.384 & 58475.301169 \\
120792 & 18.61 & 135.43493 & 17.32942 & 127.374 & 29.077 & 58475.352257 \\
474942 & 19.74 & 135.37887 & 16.29108 & 50.546 & 31.295 & 58475.302766 \\
343950 & 19.12 & 135.28006 & 16.78169 & 49.698 & 39.056 & 58475.300347 \\
373174 & 19.75 & 135.92121 & 17.04615 & 26.347 & 28.044 & 58475.300347 \\
 & 20.38 & 135.87684 & 16.88091 & 44.222 & 23.526 & 58475.300347 \\
273784 & 19.77 & 134.04942 & 17.18428 & 59.407 & 13.064 & 58475.300347 \\
190380 & 19.62 & 135.36075 & 16.11803 & 30.317 & 19.31 & 58475.300347 \\
 & 19.93 & 134.55301 & 16.34607 & 50.397 & 29.196 & 58475.300347 \\
 & 20.29 & 134.97073 & 17.45837 & 38.634 & 40.252 & 58475.300347 \\
 & 20.49 & 135.27444 & 17.38082 & 58.443 & -9.882 & 58475.300347 \\
509252 & 20.61 & 135.85848 & 16.91407 & 56.211 & 17.845 & 58475.301169 \\
350990 & 18.34 & 135.65633 & 16.96134 & 32.189 & 32.944 & 58475.300347 \\
22308 & 17.25 & 136.1153 & 17.00435 & 32.55 & 42.667 & 58475.300347 \\
456011 & 19.32 & 135.09549 & 17.39107 & 52.49 & -7.736 & 58475.300347 \\
166530 & 19.93 & 135.79449 & 17.00166 & 24.684 & 21.931 & 58475.300347 \\
 & 19.83 & 135.66423 & 16.56798 & 52.686 & 13.334 & 58475.301169 \\
 & 19.86 & 135.43679 & 17.77679 & 56.434 & -12.992 & 58475.300347 \\
266184 & 18.76 & 134.91723 & 17.22933 & 54.472 & 29.606 & 58475.300347 \\
 & 18.45 & 134.08582 & 16.68713 & 54.288 & 52.081 & 58475.300347 \\
508023 & 19.66 & 135.26764 & 17.84768 & 53.882 & 7.841 & 58475.300347 \\
 & 19.64 & 134.78689 & 17.34378 & 49.751 & -0.234 & 58475.300347 \\
 & 19.21 & 134.82559 & 16.5333 & 59.508 & -5.465 & 58475.300347 \\
178084 & 19.36 & 135.84415 & 16.56622 & 54.598 & 17.176 & 58475.300347 \\
46883 & 16.7 & 135.15826 & 17.73524 & 61.521 & 34.949 & 58475.301169 \\
 & 20.83 & 134.10219 & 17.19007 & 55.722 & -4.338 & 58475.300347 \\
 & 19.94 & 135.7474 & 16.80471 & 31.926 & 15.469 & 58475.300347 \\
250392 & 18.1 & 134.25628 & 16.49468 & 55.088 & -12.042 & 58475.300347 \\
9470 & 15.69 & 135.69139 & 17.5841 & 50.502 & 10.037 & 58475.300347 \\
 & 19.98 & 135.08763 & 17.79337 & 54.468 & -2.128 & 58475.300347 \\
5769 & 16.55 & 134.91065 & 17.49583 & 55.945 & 17.863 & 58475.300347 \\
286186 & 18.88 & 135.29339 & 17.40232 & 53.641 & 16.063 & 58475.300347 \\
 & 21.01 & 134.8931 & 18.00769 & 1.749 & -3.132 & 58475.301968 \\
 & 19.27 & 134.0956 & 16.93541 & 76.241 & -13.748 & 58475.300347 \\
17642 & 17.04 & 135.23332 & 17.75534 & 62.437 & 33.548 & 58475.300347 \\
 & 18.83 & 135.32308 & 17.94623 & 35.372 & 76.723 & 58475.300347 \\
220307 & 18.17 & 136.09491 & 16.98661 & 36.44 & -4.09 & 58475.300347 \\
 & 17.21 & 135.54982 & 17.02742 & 46.318 & 19.634 & 58475.300347 \\
 & 20.82 & 135.59494 & 17.30234 & 61.552 & 9.878 & 58475.300347 \\
300231 & 18.92 & 135.28969 & 16.74339 & 57.323 & 8.737 & 58475.300347 \\
 & 20.08 & 135.87517 & 16.59475 & 59.324 & -24.808 & 58475.301169 \\
 & 20.28 & 135.12598 & 17.20985 & 50.76 & 30.722 & 58475.300347 \\
 & 20.29 & 135.39359 & 16.3298 & 43.382 & 12.798 & 58475.300347 \\
278673 & 19.63 & 134.79528 & 16.03467 & 50.989 & 17.352 & 58475.301169 \\
188669 & 19.53 & 135.40357 & 18.02587 & 55.071 & 22.158 & 58475.300347 \\
16776 & 15.68 & 134.57762 & 16.06845 & 33.971 & 12.496 & 58475.302766 \\
 & 20.1 & 134.0741 & 16.60081 & 40.609 & 29.66 & 58475.302766 \\
 & 20.82 & 135.34161 & 16.52361 & -1.294 & 3.586 & 58475.300347 \\
261394 & 20.37 & 135.60673 & 16.52606 & 49.005 & 36.634 & 58475.300347 \\
 & 19.4 & 135.94739 & 17.5561 & 41.569 & 42.232 & 58475.300347 \\
 & 20.44 & 134.97392 & 17.00893 & 28.638 & 24.57 & 58475.300347 \\
77266 & 17.31 & 135.12389 & 16.70643 & 40.632 & 19.616 & 58475.300347 \\
542227 & 20.58 & 135.18455 & 17.41052 & 57.822 & 30.258 & 58475.301968 \\
 & 20.77 & 135.23566 & 17.24622 & 70.585 & 7.51 & 58475.301169 \\
165159 & 18.23 & 134.37372 & 17.14882 & 63.925 & 47.412 & 58475.300347 \\
 & 20.03 & 134.51699 & 17.70129 & 63.985 & -39.287 & 58475.301169 \\
 & 20.36 & 134.81407 & 17.63194 & 46.492 & 0.594 & 58475.300347 \\
474027 & 18.87 & 134.40392 & 16.64714 & 52.944 & 42.167 & 58475.301968 \\
418129 & 18.19 & 134.87483 & 17.88348 & 39.931 & -18.148 & 58475.300347 \\
6538 & 17.25 & 134.7705 & 17.12793 & 66.588 & 15.804 & 58475.300347 \\
 & 20.6 & 134.48707 & 17.02795 & 60.734 & 31.702 & 58475.300347 \\
 & 19.36 & 135.37545 & 16.18635 & 45.609 & 35.316 & 58475.300347 \\
260492 & 17.37 & 135.49424 & 16.2389 & 40.585 & -18.641 & 58475.300347 \\
 & 19.7 & 135.34271 & 16.45115 & 51.413 & -9.634 & 58475.300347 \\
4443 & 15.51 & 134.47003 & 16.55552 & 60.06 & 39.683 & 58475.300347 \\
488781 & 20.44 & 134.31472 & 16.90327 & 59.865 & 29.275 & 58475.300347 \\
103530 & 18.55 & 135.73 & 17.06041 & 54.742 & 14.324 & 58475.300347 \\
119376 & 17.49 & 135.3276 & 16.49491 & 41.236 & 1.555 & 58475.300347 \\
361051 & 18.16 & 134.62338 & 17.21597 & 33.166 & 48.899 & 58475.300347 \\
 & 20.78 & 135.31916 & 16.59105 & 54.508 & 4.007 & 58475.301169 \\
461884 & 19.89 & 135.03417 & 17.74525 & 60.311 & 24.088 & 58475.301169 \\
 & 20.66 & 135.09864 & 17.62182 & 37.968 & -9.104 & 58475.300347 \\
 & 20.45 & 134.91923 & 17.66643 & 72.066 & 20.606 & 58475.300347 \\
539738 & 18.64 & 134.89698 & 17.12403 & 41.935 & 43.034 & 58475.300347 \\
 & 20.66 & 133.94558 & 17.13093 & 58.313 & 22.471 & 58475.300347 \\
 & 19.91 & 135.1765 & 17.64411 & 35.298 & 27.18 & 58475.300347 \\
64015 & 17.44 & 134.87733 & 17.90454 & 54.156 & 12.55 & 58475.301169 \\
8978 & 16.0 & 135.14704 & 17.89014 & 65.675 & 26.978 & 58475.300347 \\
423840 & 19.85 & 134.46646 & 16.46711 & 71.194 & 29.585 & 58475.300347 \\
 & 20.22 & 134.60573 & 17.11111 & 44.773 & 34.639 & 58475.300347 \\
147773 & 19.53 & 134.80763 & 17.06768 & 63.33 & 19.264 & 58475.300347 \\
 & 20.75 & 135.46048 & 16.25968 & 57.169 & 19.055 & 58475.301169 \\
661526 & 19.93 & 135.77553 & 16.96214 & 57.715 & 1.17 & 58475.301169 \\
358445 & 19.35 & 135.23542 & 16.63116 & 47.188 & 11.246 & 58475.300347 \\
24761 & 15.67 & 135.30375 & 17.65341 & 56.99 & -18.551 & 58475.300347 \\
126301 & 17.2 & 134.16922 & 17.22408 & 40.864 & 21.326 & 58475.300347 \\
 & 20.56 & 134.37262 & 17.00279 & 78.943 & -1.699 & 58475.300347 \\
 & 20.61 & 135.50253 & 17.51833 & 82.856 & -17.575 & 58475.300347 \\
 & 17.81 & 135.91577 & 16.68033 & 85.789 & -28.674 & 58475.300347 \\
376906 & 18.83 & 134.75587 & 16.76423 & 83.931 & -52.499 & 58475.312384 \\
 & 19.88 & 135.55198 & 16.8105 & 43.921 & 9.85 & 58475.300347 \\
183688 & 18.34 & 135.57588 & 16.29158 & 58.884 & 24.35 & 58475.300347 \\
287010 & 18.24 & 134.90166 & 17.88768 & 53.569 & -3.884 & 58475.300347 \\
 & 20.19 & 135.13612 & 17.81496 & 52.154 & 26.939 & 58475.300347 \\
 & 20.17 & 135.3928 & 16.60968 & 43.422 & 16.607 & 58475.300347 \\
602145 & 19.84 & 134.38636 & 16.55467 & 58.335 & 28.688 & 58475.300347 \\
 & 20.73 & 134.83023 & 18.0493 & 38.884 & -54.864 & 58475.300347 \\
 & 19.95 & 134.73546 & 16.37383 & 54.449 & 3.078 & 58475.300347 \\
 & 18.77 & 135.76502 & 17.00668 & 54.048 & 16.711 & 58475.300347 \\
255691 & 19.85 & 134.25897 & 16.61406 & 71.802 & 19.865 & 58475.300347 \\
294239 & 19.51 & 134.39383 & 16.54091 & 60.448 & 14.357 & 58475.300347 \\
 & 18.58 & 134.57582 & 16.88552 & 51.52 & -22.064 & 58475.300347 \\
46837 & 17.9 & 135.46872 & 17.51203 & 67.994 & 21.485 & 58475.300347 \\
 & 20.15 & 134.31936 & 16.90309 & 68.194 & 26.618 & 58475.300347 \\
 & 20.4 & 134.5761 & 17.55046 & 59.958 & -0.076 & 58475.300347 \\
581272 & 18.83 & 135.13438 & 16.20795 & 55.833 & -15.7 & 58475.301169 \\
177858 & 19.63 & 134.60746 & 17.35789 & 60.952 & 10.915 & 58475.300347 \\
150968 & 18.05 & 134.51746 & 17.735 & 50.031 & 18.05 & 58475.301169 \\
18668 & 17.46 & 134.78653 & 16.97594 & 51.23 & 22.698 & 58475.301169 \\
217833 & 19.8 & 135.83334 & 16.68884 & 61.384 & 27.943 & 58475.300347 \\
663839 & 20.47 & 135.7013 & 17.22348 & 60.729 & -3.262 & 58475.300347 \\
 & 20.87 & 135.61127 & 17.78009 & 60.101 & -15.649 & 58475.300347 \\
417541 & 18.99 & 134.55183 & 17.44795 & 59.349 & 29.804 & 58475.300347 \\
 & 19.38 & 135.57666 & 16.70859 & 43.738 & 30.967 & 58475.300347 \\
 & 20.74 & 134.94554 & 17.95619 & 44.185 & 28.296 & 58475.300347 \\
214491 & 19.52 & 134.43224 & 16.85051 & 56.798 & 19.332 & 58475.300347 \\
 & 20.37 & 135.2321 & 16.33634 & 39.604 & 20.264 & 58475.300347 \\
260657 & 19.25 & 135.03231 & 17.36967 & 55.331 & 15.451 & 58475.300347 \\
 & 20.5 & 134.76348 & 16.10563 & 77.295 & -12.722 & 58475.300347 \\
552209 & 19.43 & 134.60418 & 16.19545 & 57.644 & 5.886 & 58475.300347 \\
27019 & 15.41 & 134.09866 & 17.39066 & 52.487 & 13.752 & 58475.300347 \\
 & 19.99 & 134.52439 & 17.04456 & 51.731 & 12.791 & 58475.300347 \\
198537 & 18.37 & 134.27491 & 16.53696 & 62.309 & 7.164 & 58475.300347 \\
 & 19.72 & 135.91934 & 17.32968 & 71.835 & -0.475 & 58475.300347 \\
 & 20.35 & 135.35328 & 17.17408 & 51.372 & 24.916 & 58475.300347 \\
 & 17.75 & 135.362 & 16.11228 & 67.688 & -0.616 & 58475.300347 \\
205682 & 17.99 & 136.1518 & 16.90252 & 46.979 & -14.569 & 58475.301169 \\
 & 20.32 & 135.19088 & 17.63654 & 53.788 & 4.147 & 58475.300347 \\
 & 20.7 & 134.11844 & 17.13383 & 73.308 & -7.337 & 58475.300347 \\
677931 & 18.88 & 135.46847 & 17.37129 & 39.337 & 47.97 & 58475.300347 \\
262895 & 17.92 & 134.8744 & 17.51558 & 44.239 & -3.668 & 58475.300347 \\
521280 & 20.02 & 135.64237 & 17.05015 & 49.957 & 14.994 & 58475.300347 \\
 & 17.43 & 135.02616 & 17.09924 & 62.297 & 14.274 & 58475.300347 \\
677518 & 19.1 & 134.51064 & 17.13048 & 47.823 & 25.56 & 58475.300347 \\
534631 & 20.29 & 134.26493 & 17.19806 & 63.942 & 16.999 & 58475.300347 \\
9889 & 16.25 & 135.67274 & 17.36914 & 61.378 & 23.71 & 58475.300347 \\
579476 & 19.58 & 134.56231 & 16.97512 & 27.566 & 26.953 & 58475.301968 \\
 & 18.63 & 134.62099 & 16.93608 & 21.91 & 33.829 & 58475.300347 \\
427745 & 19.54 & 134.76143 & 16.15773 & 70.677 & -4.054 & 58475.300347 \\
 & 21.02 & 134.85982 & 17.50874 & 59.171 & -2.113 & 58475.301968 \\
 & 19.78 & 134.83798 & 17.27431 & 33.599 & 25.769 & 58475.300347 \\
513029 & 18.79 & 135.00649 & 16.34023 & 60.03 & -3.812 & 58475.300347 \\
 & 19.72 & 135.12797 & 17.79551 & 57.816 & 17.039 & 58475.300347 \\
230642 & 18.52 & 135.44288 & 16.3727 & 68.2 & 5.422 & 58475.301169 \\
 & 20.57 & 135.25256 & 17.27381 & 49.12 & 17.485 & 58475.300347 \\
 & 19.8 & 135.93276 & 16.62426 & 72.478 & -14.119 & 58475.300347 \\
391646 & 19.7 & 134.94115 & 16.85069 & 53.38 & 17.932 & 58475.300347 \\
256708 & 17.67 & 134.74153 & 17.02885 & 18.433 & 25.373 & 58475.300347 \\
525327 & 19.26 & 134.82291 & 17.70847 & 63.129 & -7.61 & 58475.300347 \\
430234 & 18.66 & 134.99828 & 17.53782 & 76.034 & -13.327 & 58475.300347 \\
673147 & 19.63 & 135.8769 & 16.82667 & 64.148 & 2.995 & 58475.300347 \\
130163 & 19.15 & 134.96891 & 16.84801 & 59.807 & 25.06 & 58475.300347 \\
 & 19.74 & 134.84806 & 16.93507 & 55.956 & 31.91 & 58475.300347 \\
538702 & 19.58 & 135.55464 & 16.4501 & 56.813 & 23.288 & 58475.300347 \\
415273 & 19.81 & 135.8166 & 17.55665 & 51.653 & 10.472 & 58475.300347 \\
148710 & 18.53 & 134.72185 & 16.12625 & 64.92 & 13.165 & 58475.300347 \\
 & 19.85 & 135.58038 & 16.53938 & 49.747 & 13.061 & 58475.300347 \\
 & 18.35 & 135.51165 & 17.31583 & 49.308 & -9.374 & 58475.300347 \\
 & 19.51 & 134.79216 & 17.22858 & 49.39 & 33.797 & 58475.300347 \\
 & 20.82 & 134.09934 & 17.1773 & 60.706 & -2.61 & 58475.300347 \\
467952 & 20.12 & 134.57006 & 17.41323 & 60.037 & 30.337 & 58475.301169 \\
 & 20.14 & 134.51405 & 16.12569 & 49.237 & 5.602 & 58475.302766 \\
 & 20.37 & 135.76149 & 17.19018 & 53.882 & 9.738 & 58475.300347 \\
498850 & 19.19 & 135.54189 & 16.95697 & 46.745 & 24.606 & 58475.301968 \\
621143 & 19.84 & 134.96609 & 16.8278 & 65.547 & 27.659 & 58475.300347 \\
 & 17.8 & 134.82962 & 17.84133 & 56.016 & 14.378 & 58475.300347 \\
 & 19.66 & 135.2296 & 16.53891 & 53.674 & 20.596 & 58475.300347 \\
 & 20.35 & 134.10112 & 17.13622 & 56.182 & 13.122 & 58475.300347 \\
 & 20.54 & 135.03577 & 17.10945 & 62.049 & -5.566 & 58475.301169 \\
 & 18.41 & 135.43409 & 17.3294 & 46.147 & 17.276 & 58475.345914 \\
600245 & 20.22 & 134.64215 & 16.98929 & 57.779 & 13.284 & 58475.300347 \\
318146 & 19.28 & 134.52753 & 17.43151 & 63.579 & 3.269 & 58475.300347 \\
95582 & 17.56 & 134.12376 & 16.6841 & 53.13 & 38.164 & 58475.300347 \\
44014 & 16.97 & 135.91186 & 17.26817 & 41.452 & 23.414 & 58475.300347 \\
 & 20.77 & 135.05198 & 16.69015 & 36.056 & 23.54 & 58475.301968 \\
77420 & 16.21 & 135.21516 & 16.4483 & 28.509 & -19.984 & 58475.302766 \\
 & 20.38 & 134.65146 & 17.70069 & 46.886 & 34.067 & 58475.300347 \\
463401 & 19.72 & 134.70812 & 17.87152 & 43.682 & 36.565 & 58475.300347 \\
 & 20.42 & 135.24513 & 17.59055 & 57.772 & 5.267 & 58475.300347 \\
472892 & 19.92 & 135.66265 & 16.61926 & 55.977 & 29.243 & 58475.300347 \\
 & 19.24 & 135.55943 & 17.17658 & 49.82 & 28.397 & 58475.300347 \\
255808 & 17.76 & 135.54513 & 17.41992 & 39.632 & 33.862 & 58475.300347 \\
 & 20.69 & 134.93874 & 18.00672 & 43.501 & 24.703 & 58475.300347 \\
216107 & 20.21 & 134.59559 & 17.7875 & 66.563 & 21.114 & 58475.300347 \\
291327 & 19.23 & 133.99728 & 17.00698 & 64.214 & 10.35 & 58475.300347 \\
431811 & 19.64 & 135.31303 & 16.75942 & 57.649 & 14.828 & 58475.301169 \\
 & 20.17 & 134.64992 & 17.1545 & 45.905 & 75.348 & 58475.300347 \\
674128 & 18.12 & 134.62777 & 16.92916 & 61.382 & -43.229 & 58475.300347 \\
89964 & 17.36 & 135.07826 & 17.18778 & 47.276 & 23.893 & 58475.300347 \\
 & 19.12 & 135.24719 & 16.53393 & 34.629 & 18.644 & 58475.300347 \\
 & 19.75 & 134.94336 & 17.3722 & 48.585 & 24.016 & 58475.300347 \\
 & 19.29 & 134.52739 & 17.42166 & 60.691 & 26.028 & 58475.300347 \\
499477 & 20.02 & 135.57338 & 16.77963 & 63.713 & 39.11 & 58475.300347 \\
 & 20.5 & 134.64071 & 16.93852 & 34.507 & 12.726 & 58475.300347 \\
170158 & 18.7 & 134.56489 & 17.10633 & 50.193 & 20.329 & 58475.302766 \\
 & 20.13 & 134.53876 & 16.29287 & 26.634 & 35.057 & 58475.300347 \\
 & 20.1 & 134.38619 & 16.59145 & 47.522 & 14.101 & 58475.301169 \\
41901 & 16.72 & 134.32881 & 17.61989 & 43.198 & 18.792 & 58475.305961 \\
381522 & 19.38 & 135.49721 & 17.27391 & 60.77 & 7.276 & 58475.300347 \\
100533 & 17.95 & 135.11626 & 16.73943 & 65.071 & 21.946 & 58475.300347 \\
 & 19.58 & 134.66398 & 17.37553 & 27.568 & 23.173 & 58475.300347 \\
 & 20.27 & 135.53247 & 16.92004 & 53.636 & -2.405 & 58475.301169 \\
352472 & 18.07 & 134.05337 & 17.12481 & 38.357 & 72.414 & 58475.300347 \\
256636 & 18.28 & 135.36548 & 17.93669 & 60.883 & 4.316 & 58475.300347 \\
408797 & 20.0 & 134.49007 & 17.34945 & 79.785 & 17.413 & 58475.300347 \\
 & 19.76 & 134.2307 & 16.55724 & 37.475 & 52.787 & 58475.301169 \\
 & 20.47 & 133.97791 & 17.15478 & 47.449 & -21.524 & 58475.300347 \\
 & 20.31 & 135.3578 & 16.65846 & 52.213 & 29.905 & 58475.300347 \\
351861 & 20.18 & 134.71354 & 17.27571 & 62.581 & 17.24 & 58475.300347 \\
 & 18.85 & 134.27187 & 16.81163 & 44.321 & 54.972 & 58475.300347 \\
508599 & 20.03 & 135.01639 & 17.48889 & 75.36 & 8.471 & 58475.300347 \\
 & 20.44 & 134.61301 & 17.57297 & 53.024 & 21.593 & 58475.301169 \\
 & 20.05 & 135.73037 & 17.10609 & 41.069 & 26.773 & 58475.300347 \\
215060 & 18.99 & 134.82543 & 16.45099 & 52.328 & 21.262 & 58475.300347 \\
 & 20.11 & 135.3333 & 16.23553 & 55.839 & 12.704 & 58475.300347 \\
 & 19.94 & 134.86455 & 17.59219 & 46.684 & 35.15 & 58475.300347 \\
 & 20.31 & 134.91288 & 17.05437 & 28.373 & 24.613 & 58475.300347 \\
 & 19.81 & 134.56303 & 16.49217 & 43.829 & 14.022 & 58475.301169 \\
381436 & 20.05 & 135.22639 & 17.61307 & 64.614 & 19.49 & 58475.300347 \\
72997 & 16.69 & 134.86342 & 16.48966 & 51.379 & -7.078 & 58475.300347 \\
107889 & 16.93 & 135.51453 & 17.00693 & 33.066 & 48.546 & 58475.300347 \\
347455 & 19.3 & 135.24629 & 17.80423 & 51.194 & 25.297 & 58475.300347 \\
 & 20.59 & 134.27526 & 16.73583 & 32.724 & 0.14 & 58475.300347 \\
  \enddata
  \tablenotetext{a}{$\alpha$ and $\delta$ refer to right ascension and declination, respectively.}
  \tablenotetext{b}{Observation time refers to the time at which the R.A. and Dec. values were observed. These were the first exposure the detected trajectory intersected.}
  \end{deluxetable*}

\startlongtable
\begin{deluxetable*}{ccccccc}
\tablenum{3}
\tablecaption{Night Two Detections \label{tab:night2}}
\tablewidth{0pt}
\tablehead{
\colhead{MPC Designation} & \colhead{$m_r$} & \colhead{RA} & \colhead{Dec} & \colhead{$\pi_{RA}\cos{\delta}$} & \colhead{$\pi_{\delta}$} & \colhead{Observation Time}  \\ 
\colhead{\,} & \colhead{magnitudes} & \colhead{$^\circ$} & \colhead{$^{\circ}$} & \colhead{$\arcsec \, hr^{-1}$} & \colhead{$\arcsec \, hr^{-1}$} & \colhead{MJD}}
\decimalcolnumbers
\startdata
170608 & 18.63 & 135.22188 & 17.36393 & 55.573 & 15.898 & 58476.300833 \\
576938 & 18.74 & 136.43999 & 15.95802 & 44.19 & 20.596 & 58476.302454 \\
 & 20.06 & 135.03027 & 16.6883 & 36.68 & 28.919 & 58476.300833 \\
 & 19.86 & 135.8096 & 17.08086 & 32.103 & 19.102 & 58476.300833 \\
277630 & 18.92 & 135.66741 & 16.22197 & 42.883 & -1.184 & 58476.300833 \\
341429 & 18.58 & 135.79427 & 15.79003 & 47.4 & 23.281 & 58476.302454 \\
215444 & 19.65 & 136.99119 & 17.07061 & 63.862 & 33.106 & 58476.301655 \\
522895 & 19.52 & 135.74501 & 17.6383 & 44.641 & 24.624 & 58476.300833 \\
 & 19.59 & 135.69084 & 16.83357 & 35.794 & 17.402 & 58476.300833 \\
299971 & 18.07 & 135.65043 & 16.15148 & 56.851 & -9.274 & 58476.300833 \\
7059 & 15.24 & 136.86659 & 17.06729 & 48.982 & 8.442 & 58476.300833 \\
358445 & 19.32 & 135.14534 & 16.64968 & 50.581 & 12.301 & 58476.300833 \\
111700 & 17.36 & 135.93196 & 16.55727 & 42.168 & 29.754 & 58476.300833 \\
350990 & 18.88 & 135.59726 & 17.01963 & 35.084 & 34.42 & 58476.300833 \\
460720 & 19.32 & 135.7289 & 16.15551 & 73.714 & 16.448 & 58476.300833 \\
177937 & 18.81 & 135.68836 & 15.88187 & 54.286 & 13.468 & 58476.300833 \\
67077 & 17.41 & 136.32295 & 17.19308 & 69.773 & 7.776 & 58476.300833 \\
220480 & 19.3 & 136.2989 & 16.11685 & 59.96 & 25.11 & 58476.302454 \\
214124 & 17.65 & 135.74019 & 17.01302 & 36.728 & 43.888 & 58476.300833 \\
330131 & 19.43 & 136.31937 & 15.79908 & 63.384 & 30.625 & 58476.301655 \\
 & 20.65 & 136.67627 & 16.49828 & 49.039 & 18.896 & 58476.300833 \\
100878 & 17.78 & 135.07358 & 16.67615 & 44.532 & 27.266 & 58476.300833 \\
10054 & 16.18 & 136.33663 & 16.61992 & 52.22 & 16.78 & 58476.302454 \\
118468 & 17.42 & 136.51153 & 16.62008 & 46.452 & 26.827 & 58476.302454 \\
 & 20.27 & 136.34698 & 16.40979 & 69.765 & 5.317 & 58476.302454 \\
198563 & 18.3 & 136.33217 & 16.34901 & 53.015 & 13.626 & 58476.300833 \\
 & 18.89 & 135.41844 & 17.29914 & 52.348 & -8.597 & 58476.324873 \\
286186 & 19.03 & 135.19343 & 17.42729 & 56.234 & 15.854 & 58476.308854 \\
 & 20.41 & 135.31207 & 16.63987 & 46.599 & 19.39 & 58476.300833 \\
120792 & 17.65 & 135.34902 & 17.37105 & 50.908 & 26.273 & 58476.300833 \\
620113 & 20.35 & 136.33872 & 17.04476 & 77.986 & 18.418 & 58476.300833 \\
 & 19.23 & 135.28843 & 16.83695 & 38.147 & -1.303 & 58476.300833 \\
77266 & 17.47 & 135.05051 & 16.73986 & 45.313 & 20.725 & 58476.300833 \\
 & 20.07 & 135.47953 & 17.31833 & 65.547 & 10.332 & 58476.300833 \\
99083 & 17.01 & 135.86024 & 16.56101 & 50.211 & 0.515 & 58476.300833 \\
 & 19.98 & 135.75424 & 17.04098 & 29.064 & 23.256 & 58476.301655 \\
308315 & 19.91 & 135.33333 & 16.65702 & 63.726 & 9.115 & 58476.300833 \\
12489 & 17.32 & 136.49334 & 16.071 & 45.203 & 14.713 & 58476.300833 \\
 & 19.78 & 135.24967 & 16.43344 & 54.615 & -8.489 & 58476.301655 \\
673147 & 19.65 & 135.75638 & 16.83001 & 68.017 & 2.912 & 58476.312106 \\
261394 & 19.88 & 135.48795 & 16.56176 & 52.07 & 13.334 & 58476.301655 \\
410328 & 18.68 & 136.00154 & 16.86384 & 40.932 & 21.29 & 58476.300833 \\
563034 & 18.95 & 135.32085 & 17.38142 & 54.671 & 49.993 & 58476.300833 \\
77422 & 17.33 & 136.19751 & 17.23524 & 48.395 & 9.425 & 58476.300833 \\
504540 & 19.76 & 136.40104 & 17.25066 & 56.621 & 11.898 & 58476.300833 \\
175458 & 16.44 & 136.52869 & 16.5626 & 46.28 & 26.737 & 58476.300833 \\
 & 20.21 & 135.76753 & 16.55105 & 63.337 & -23.436 & 58476.300833 \\
272748 & 18.42 & 136.03915 & 16.42488 & 48.578 & 24.275 & 58476.300833 \\
11775 & 17.05 & 136.25558 & 16.4975 & 56.937 & 16.153 & 58476.300833 \\
166603 & 18.37 & 136.61049 & 16.28888 & 57.769 & 15.102 & 58476.302454 \\
601077 & 19.53 & 136.4359 & 17.2971 & 54.517 & -5.072 & 58476.301655 \\
80806 & 17.07 & 135.86608 & 17.7458 & 48.77 & 37.85 & 58476.300833 \\
369867 & 19.16 & 135.34519 & 17.01431 & 66.979 & 34.506 & 58476.300833 \\
263991 & 17.77 & 135.71841 & 15.84891 & 58.631 & 38.516 & 58476.300833 \\
193614 & 17.7 & 136.23192 & 17.71814 & 39.663 & 49.021 & 58476.300833 \\
753413 & 19.79 & 136.6366 & 16.59783 & 29.121 & 2.257 & 58476.300833 \\
 & 19.17 & 136.66966 & 16.83054 & 54.65 & 1.386 & 58476.300833 \\
 & 19.89 & 136.34304 & 16.71795 & 33.92 & 30.262 & 58476.300833 \\
601979 & 18.24 & 136.14168 & 17.22774 & 55.841 & -9.842 & 58476.300833 \\
233849 & 19.24 & 135.88851 & 16.95816 & 52.537 & 17.359 & 58476.300833 \\
 & 19.75 & 136.66689 & 17.45598 & 53.062 & 26.924 & 58476.301655 \\
 & 19.39 & 135.46489 & 17.22491 & 52.28 & 28.865 & 58476.300833 \\
 & 20.23 & 135.65556 & 17.15474 & 44.422 & 29.509 & 58476.300833 \\
429278 & 19.42 & 135.62698 & 17.22858 & 51.532 & 29.027 & 58476.300833 \\
308359 & 18.63 & 135.80359 & 17.02565 & 81.777 & -20.815 & 58476.300833 \\
 & 20.18 & 135.78246 & 16.0388 & 63.717 & 5.868 & 58476.300833 \\
127602 & 17.34 & 136.2752 & 16.21634 & 70.532 & 2.218 & 58476.300833 \\
538702 & 19.71 & 135.44846 & 16.48989 & 59.442 & 23.436 & 58476.300833 \\
532854 & 19.76 & 136.79087 & 16.20461 & 71.127 & -6.962 & 58476.301655 \\
255808 & 17.72 & 135.4741 & 17.47971 & 45.041 & 35.759 & 58476.300833 \\
343388 & 18.5 & 135.70054 & 17.6587 & 68.257 & -11.092 & 58476.300833 \\
656468 & 20.32 & 136.74881 & 17.0585 & 52.987 & 8.986 & 58476.300833 \\
594360 & 20.5 & 136.7577 & 16.44631 & 51.739 & 24.66 & 58476.302454 \\
277377 & 19.67 & 136.32328 & 17.12817 & 62.294 & 22.669 & 58476.300833 \\
498850 & 19.13 & 135.4559 & 16.99877 & 50.57 & 25.117 & 58476.300833 \\
166030 & 18.15 & 135.64987 & 17.23355 & 47.223 & 21.514 & 58476.300833 \\
473267 & 18.87 & 135.81728 & 17.32215 & 42.399 & 49.874 & 58476.300833 \\
 & 20.08 & 136.31433 & 16.6182 & 50.547 & -0.079 & 58476.302454 \\
119376 & 17.26 & 135.25554 & 16.49822 & 43.99 & 2.772 & 58476.300833 \\
 & 19.73 & 135.61529 & 17.42385 & 51.244 & 14.9 & 58476.300833 \\
431072 & 19.47 & 136.97429 & 17.05782 & 43.853 & 0.713 & 58476.301655 \\
626709 & 19.15 & 135.86705 & 16.40152 & 59.217 & 4.345 & 58476.301655 \\
663839 & 20.13 & 135.58638 & 17.21575 & 63.547 & -3.197 & 58476.300833 \\
510285 & 19.98 & 135.32888 & 16.52914 & 67.827 & 5.166 & 58476.301655 \\
365075 & 18.91 & 136.48842 & 16.15177 & 59.528 & 10.94 & 58476.322488 \\
386506 & 19.27 & 136.09069 & 17.75528 & 46.456 & 27.302 & 58476.300833 \\
16500 & 17.23 & 136.64005 & 16.51016 & 53.986 & 22.626 & 58476.300833 \\
 & 19.8 & 135.89591 & 16.90669 & 61.6 & 8.903 & 58476.300833 \\
147181 & 18.21 & 136.27812 & 17.51217 & 65.831 & 17.762 & 58476.300833 \\
324998 & 18.25 & 136.69835 & 16.6882 & 25.028 & -32.699 & 58476.300833 \\
 & 19.36 & 136.24379 & 16.13965 & 52.325 & 22.198 & 58476.300833 \\
348902 & 20.05 & 136.93307 & 16.53251 & 39.312 & -20.286 & 58476.302454 \\
491052 & 18.58 & 135.972 & 16.36937 & 55.269 & 32.659 & 58476.300833 \\
42019 & 17.67 & 135.47326 & 17.05354 & 58.027 & 18.158 & 58476.300833 \\
68050 & 18.01 & 136.77643 & 16.49843 & 76.136 & 14.782 & 58476.300833 \\
160942 & 17.51 & 135.98831 & 17.58401 & 71.649 & 5.44 & 58476.300833 \\
9889 & 15.79 & 135.55901 & 17.40975 & 66.005 & 25.11 & 58476.300833 \\
 & 20.12 & 135.56822 & 16.58966 & 54.599 & 13.216 & 58476.300833 \\
503442 & 19.55 & 134.97665 & 16.81495 & 59.507 & 5.252 & 58476.300833 \\
343950 & 18.84 & 135.18567 & 16.84944 & 53.914 & 39.654 & 58476.300833 \\
 & 20.44 & 136.15398 & 17.74642 & 32.686 & 6.793 & 58476.300833 \\
205682 & 18.07 & 136.06945 & 16.87638 & 52.201 & -12.812 & 58476.300833 \\
 & 19.63 & 135.80239 & 17.77756 & -2.331 & 3.362 & 58476.308854 \\
 & 20.6 & 136.59895 & 16.5701 & 70.376 & -17.874 & 58476.301655 \\
 & 18.86 & 136.24951 & 17.03423 & 67.871 & -12.499 & 58476.300833 \\
211948 & 18.42 & 136.31032 & 17.33449 & 33.743 & 10.382 & 58476.300833 \\
36186 & 16.65 & 136.68077 & 16.10726 & 49.262 & 27.374 & 58476.302454 \\
 & 20.36 & 136.17558 & 16.43811 & 57.801 & 17.737 & 58476.300833 \\
294715 & 18.65 & 136.76117 & 16.21089 & 61.698 & -9.374 & 58476.301655 \\
214666 & 17.7 & 135.49641 & 16.61914 & 37.035 & 69.217 & 58476.302454 \\
432453 & 19.17 & 135.68294 & 15.83491 & 47.355 & 15.379 & 58476.300833 \\
 & 17.95 & 136.48921 & 16.15486 & 45.267 & -23.018 & 58476.300833 \\
69945 & 18.43 & 135.7733 & 15.97912 & 62.584 & 14.483 & 58476.300833 \\
327173 & 18.5 & 136.99326 & 16.56207 & 5.321 & 31.874 & 58476.302454 \\
 & 20.5 & 135.6533 & 16.89193 & 51.873 & 13.039 & 58476.300833 \\
9470 & 15.69 & 135.59699 & 17.59959 & 54.588 & 10.469 & 58476.302454 \\
 & 19.45 & 135.93277 & 15.99409 & 51.228 & -27.666 & 58476.300833 \\
215249 & 18.56 & 136.21533 & 16.1978 & 67.227 & 40.871 & 58476.300833 \\
74196 & 17.84 & 136.56215 & 16.23027 & 54.267 & 26.165 & 58476.300833 \\
 & 19.48 & 135.29135 & 16.24837 & 48.809 & 36.925 & 58476.300833 \\
 & 19.89 & 136.13554 & 16.94882 & 57.722 & 8.942 & 58476.302454 \\
160952 & 18.12 & 136.53272 & 16.7094 & 74.49 & 31.54 & 58476.300833 \\
 & 19.42 & 136.61242 & 17.31761 & 34.842 & -25.805 & 58476.300833 \\
 & 20.26 & 136.63788 & 16.00316 & 60.001 & -7.2 & 58476.301655 \\
368106 & 18.9 & 136.89205 & 16.46623 & 49.552 & 26.496 & 58476.300833 \\
 & 20.27 & 135.29237 & 17.13186 & 49.182 & -30.974 & 58476.301655 \\
 & 18.5 & 136.82018 & 16.52299 & 51.701 & -8.705 & 58476.3161 \\
661526 & 20.16 & 135.66744 & 16.96293 & 60.218 & 2.531 & 58476.300833 \\
287521 & 18.31 & 135.54068 & 17.71363 & 43.237 & 44.68 & 58476.300833 \\
230642 & 18.65 & 135.31528 & 16.38053 & 72.082 & 6.854 & 58476.300833 \\
97240 & 16.73 & 136.01248 & 16.00087 & 47.499 & 6.235 & 58476.300833 \\
474942 & 20.2 & 135.28387 & 16.34534 & 54.115 & 33.091 & 58476.300833 \\
 & 20.08 & 135.84387 & 16.13753 & 44.925 & 21.29 & 58476.301655 \\
 & 19.16 & 136.55238 & 17.0633 & 29.329 & 49.28 & 58476.300833 \\
180367 & 18.35 & 135.65373 & 16.72844 & 60.234 & 32.346 & 58476.300833 \\
356324 & 19.36 & 136.43649 & 16.72724 & 46.695 & 44.309 & 58476.301655 \\
210410 & 19.18 & 136.1321 & 16.63173 & 49.737 & 13.712 & 58476.300833 \\
334595 & 19.05 & 135.61312 & 17.67213 & 57.434 & 17.716 & 58476.300833 \\
 & 19.59 & 135.13039 & 16.57398 & 56.332 & 21.038 & 58476.300833 \\
 & 20.35 & 136.64475 & 17.03865 & 50.759 & -7.153 & 58476.300833 \\
 & 20.27 & 136.11317 & 16.54213 & 49.128 & 20.527 & 58476.300833 \\
 & 20.45 & 136.29677 & 17.3111 & 54.077 & 18.58 & 58476.300833 \\
171567 & 19.05 & 135.69274 & 16.19179 & 71.332 & 4.918 & 58476.300833 \\
564072 & 19.46 & 136.26338 & 15.95012 & 61.111 & -16.308 & 58476.302454 \\
196982 & 17.73 & 136.16396 & 16.57139 & 49.359 & 35.86 & 58476.300833 \\
499477 & 20.59 & 135.47242 & 16.82739 & 49.813 & 10.159 & 58476.300833 \\
 & 19.5 & 135.39774 & 16.84382 & 35.954 & 6.016 & 58476.300833 \\
103530 & 18.4 & 135.62942 & 17.08498 & 57.766 & 15.491 & 58476.300833 \\
139191 & 18.39 & 136.44185 & 16.54856 & 48.923 & 23.479 & 58476.300833 \\
310387 & 18.42 & 136.48133 & 17.72875 & 39.835 & 8.064 & 58476.301655 \\
77420 & 16.44 & 135.17018 & 16.41232 & 32.634 & -20.563 & 58476.300833 \\
 & 20.6 & 136.36311 & 16.18875 & 63.489 & 1.35 & 58476.300833 \\
 & 19.35 & 136.12582 & 16.95941 & 45.987 & 74.678 & 58476.300833 \\
 & 18.96 & 136.29627 & 15.75668 & 18.519 & 31.871 & 58476.302454 \\
 & 19.59 & 135.78388 & 17.30166 & 29.191 & 23.206 & 58476.301655 \\
208987 & 17.96 & 135.78051 & 16.51568 & 41.031 & 22.406 & 58476.300833 \\
 & 19.47 & 136.34439 & 16.55707 & 56.927 & 1.451 & 58476.300833 \\
141170 & 18.28 & 135.83898 & 16.36287 & 47.654 & 23.746 & 58476.300833 \\
 & 18.16 & 136.50655 & 16.57531 & 59.82 & 5.724 & 58476.300833 \\
210210 & 19.12 & 136.4391 & 15.99642 & 56.034 & 11.7 & 58476.300833 \\
12538 & 16.32 & 136.75815 & 16.15224 & 51.782 & 3.636 & 58476.302454 \\
268038 & 18.65 & 136.91207 & 17.19404 & 60.57 & 2.898 & 58476.300833 \\
18456 & 15.66 & 136.47852 & 17.32713 & 36.5 & 14.688 & 58476.300833 \\
326927 & 18.26 & 135.46241 & 17.5636 & 47.769 & -4.997 & 58476.300833 \\
 & 20.28 & 136.48621 & 17.23242 & 60.853 & 30.47 & 58476.301655 \\
118976 & 18.62 & 136.30441 & 17.38439 & 54.625 & 22.288 & 58476.300833 \\
83762 & 16.82 & 136.59937 & 16.05942 & 58.064 & -4.025 & 58476.301655 \\
535650 & 19.39 & 135.77725 & 16.50283 & 42.691 & 34.798 & 58476.300833 \\
 & 17.8 & 136.45865 & 16.60754 & 50.533 & 30.287 & 58476.302454 \\
300100 & 20.48 & 135.92023 & 16.89878 & 74.564 & 3.992 & 58476.300833 \\
422252 & 19.81 & 136.49046 & 16.02948 & 49.88 & 30.092 & 58476.300833 \\
525560 & 19.87 & 136.24626 & 17.75111 & 53.109 & 23.285 & 58476.300833 \\
217833 & 19.3 & 135.72047 & 16.73632 & 63.379 & 27.961 & 58476.300833 \\
62841 & 17.66 & 136.67904 & 16.17763 & 61.906 & 10.156 & 58476.300833 \\
546659 & 20.46 & 136.27346 & 16.68588 & 66.792 & 25.52 & 58476.325683 \\
 & 20.22 & 135.21984 & 16.59767 & 57.936 & 5.368 & 58476.300833 \\
135382 & 18.37 & 135.71195 & 16.19279 & 62.63 & 15.304 & 58476.300833 \\
302010 & 19.77 & 134.94177 & 16.84369 & 65.979 & 8.244 & 58476.300833 \\
 & 19.7 & 136.7685 & 17.13505 & 30.098 & 1.39 & 58476.300833 \\
227354 & 18.58 & 136.56422 & 16.77846 & 60.666 & 15.127 & 58476.302454 \\
 & 19.8 & 135.80003 & 16.59761 & 74.196 & -12.827 & 58476.300833 \\
 & 20.68 & 135.60756 & 17.34904 & 53.114 & 14.332 & 58476.300833 \\
 & 19.35 & 135.19157 & 16.87652 & 54.537 & -13.032 & 58476.300833 \\
300231 & 18.71 & 135.18482 & 16.75723 & 60.001 & 9.76 & 58476.301655 \\
677931 & 18.38 & 135.39582 & 17.45446 & 42.997 & 49.021 & 58476.300833 \\
 & 20.9 & 135.69761 & 16.20796 & 60.43 & -8.784 & 58476.300833 \\
77485 & 17.29 & 135.53661 & 16.30218 & 48.007 & 24.638 & 58476.300833 \\
 & 19.73 & 135.78325 & 17.3273 & 74.949 & 0.594 & 58476.301655 \\
 & 20.39 & 136.59471 & 16.56258 & 38.316 & 9.799 & 58476.301655 \\
 & 19.61 & 135.30874 & 16.15239 & 34.6 & 21.38 & 58476.300833 \\
 & 19.24 & 135.5803 & 17.50614 & 13.088 & 51.754 & 58476.300833 \\
  \enddata
  \end{deluxetable*}

\acknowledgments

This work was performed under the auspices of the U.S. Department of Energy by Lawrence Livermore National Laboratory under Contract DE-AC52-07NA27344 and was supported by the LLNL-LDRD Program under Projects 17-ERD-120, 19-SI-004, and 20-ER-025. This work was based on observations obtained at Cerro Tololo Inter-American Observatory a division of the National Optical Astronomy Observatories, which is operated by the Association of Universities for Research in Astronomy, Inc. under cooperative agreement with the National Science Foundation.

We thank the PI of the PALS survey (William Dawson) for granting us eight hours of observational time over four nights during which their targets were not observable. We thank Maya Gokhale, Eddie Schlafly, Will Dawson, and Josh Meyers for useful discussions. We thank the Pan-STARRS team for their dutiful photometry catalogs for calibration. The Pan-STARRS1 Surveys (PS1) and the PS1 public science archive have been made possible through contributions by the Institute for Astronomy, the University of Hawaii, the Pan-STARRS Project Office, the Max-Planck Society and its participating institutes, the Max Planck Institute for Astronomy, Heidelberg and the Max Planck Institute for Extraterrestrial Physics, Garching, The Johns Hopkins University, Durham University, the University of Edinburgh, the Queen's University Belfast, the Harvard-Smithsonian Center for Astrophysics, the Las Cumbres Observatory Global Telescope Network Incorporated, the National Central University of Taiwan, the Space Telescope Science Institute, the National Aeronautics and Space Administration under Grant No. NNX08AR22G issued through the Planetary Science Division of the NASA Science Mission Directorate, the National Science Foundation Grant No. AST-1238877, the University of Maryland, Eotvos Lorand University (ELTE), the Los Alamos National Laboratory, and the Gordon and Betty Moore Foundation. This research has made use of NASA's Astrophysics Data System Bibliographic Services. This research made use of NumPy \citep{van2011numpy}. This research made use of Astropy, a community-developed core Python package for Astronomy \citep{2018AJ....156..123A, 2013A&A...558A..33A}. This research made use of matplotlib, a Python library for publication quality graphics \citep{Hunter:2007}. This research made use of pandas \citep{McKinney_2010, McKinney_2011}. This research made use of the REBOUND integrator package \citep{reinliu2012rebound}. 

\vspace{5mm}
\facilities{CTIO(Blanco/DECam)}
\newpage
\bibliography{asteroids}{}
\bibliographystyle{aasjournal}

\end{document}